# On the correlation of earthquake occurrence among major fault zones in the eastern margin of the Tibetan Plateau by Big Data Analysis

Zili Zhou [a,e,*], Huai Zhang [a,b,*] , Huihong Cheng [a], Chunlin Deng [a], Yaolin Shi [a], Shi Chen [b], Xiwu Luan [c], David A. Yuen [d]

[a] Key Laboratory of Computational Geodynamics, University of Chinese Academy of Sciences, No. 19(A) Yuquanlu Ave., Beijing 100049, China

[b] Institute of Geophysics, China Earthquake Administration, No.5 Minzudaxue South Road, Beijing 100081, China.

[c] Laboratory for Marine Mineral Resources, Qingdao National Laboratory for Marine Science and Technology, Qingdao 266237, China

[d] Department of Earth Sciences, University of Minnesota, 310 Pillsbury Drive SE, Minneapolis, MN55455-0231, USA

[e] Sonny Astani Department of Civil & Environmental Engineering, University of Southern California, 3620 S Vermont Ave, Los Angeles CA90089, USA

* Authors of correspondence: zilizhou@usc.edu, hzhang@ucas.ac.cn

**Main contributions:**

1. Previously unrecognized strong correlation among major faults with highly earthquake risk in the eastern margin of the Tibetan Plateau is uncovered.
2. Fault pairs of far distance show high synchronicity with each other are identified.
3. Adjacent faults zones surprisingly exhibit low correlation of seismicity.
4. Seismic belts turn out to be just a group of earthquakes rather than an indicator of close internal relationship of faults.
5. The location of the area where seismicity keeps increasing comparing to every previous no-big-earthquake period might be where the next big event happens.

**ABSTRACT**

The subsequent series of responses to big events may exhibit a synchronicity of event number, frequency and energy release in different fault zones. This synchronicity is a reliable source for probing non-intuitive geological structures, assessing regional seismicity hazard map and even predicting the next big events. The synchronicity of main faults in the eastern margin of the Qinghai-Tibetan Plateau is still unknown to us. We propose to examine the correlation of earthquake occurrence among different fault zones to indicate this synchronicity and to obtain a preliminary understanding of geodynamics processes and the unrecognized characteristics of deep evolution in the eastern margin of the Qinghai-Tibetan Plateau. We estimate temporal changes of completeness level, frequency seismicity, and intensity seismicity, referring respectively to Mc, Z, and E values, of 21 main fault zones, using a seismic catalogue from 1970 to 2015. Our results reveal that six fault zone pairs of fault zones exhibit relative high correlation (>0.6) by all three indicators, while four fault zone pairs are non-adjacent with close internal affinity offsetting the limit of spatial distance, such as the pair of Rongjing-mabian fault and Minjiang-huya fault. In addition, six strike-slip pairs of faults are all either parallel or perpendicular with each other, which is coincidental with stress direction distribution within a same tectonic stress field. Most strikingly, some fault zone pairs showing typical high correlation (>0.8) of seismicity frequency or seismicity intensity (e.g. the pair of Muli fault and Huarongshan fault zones, and the pair of Fubianhe fault, Xianshuihe fault and Longmenshan fault zones), the faults surprisingly belong to neither the same seismic belt nor the same geological block, exhibiting a regional scale remote triggering pattern of earthquakes or structures. Meanwhile, in the same seismic belts or geological block, only a few faults zones show innate close correlation with each other. An embryonic pattern to predict the next possible events will also be presented. This correlation analysis discovers a previously unrecognized strong coupling relationship among main faults with high earthquake risk in the eastern margin of the Qinghai-Tibetan Plateau.
**Keywords:** Qinghai-Tibetan Plateau; faults correlation; Z test; completeness; seismicity; seismicity energy;

## 1 Introduction

Following a large earthquake, the regional seismicity pattern changes dramatically(Toda *et al.*, 2008), and the number, frequency and energy release sharply increases in some regions while decreases in some other regions (H Zhang *et al.*, 2016). The innate responses to the big events in neighbor fault zones may exhibit a simultaneous change of number, frequency and energy release pattern, whereby the two or more fault zones achieve synchronicity. The synchronicity of seismicity sequence of regional fault zones is one of the reliable constrains for probing the non-intuitive geological structure, assessing regional seismicity hazard map and even predicting the next big events (Kumar *et al.*, 2010; Salazar *et al.*, 2016; Xiang and Wang, 2017). The seismicity occurrence correlation among different fault zones is an efficient and direct indicator of the degree of synchronicity between fault zone pairs. This synchronicity can lead to a general pattern of the deep geological evolution.

Previous investigations examined the correlation between the seismicity occurrence of adjoining fault zones or special areas to evaluate the seismic hazard potential. Kumar et al. (2010) used the qualitative correlation of two far field fault zones along the Himalayan frontal thrust to examine the paleo-seismicity by determine their timing, size, and spatial distribution. Bayrak et al. (2017) obtained qualitative correlation of 15 different seismic-tectonic zones by b and DC values to understand seismicity in Western Anatolia, Turkey. Chen et al. (2017) proposed a quantitative correlation method between in-panel faults and high-stress areas to study the regional mechanism, through Monte Carlo simulation and point process statistics using fault traces and seismic velocities. In this study, we want to know that the seismicity occurrence correlation, which includes frequency seismicity and intensity seismicity, among different fault zones in the eastern margin of the Qinghai-Tibet Plateau, aiming to understand the earthquake occurrence and possible geodynamics processes behind earthquake triggering mechanism in this area. According to the previous main earthquakes distributed in different fault zones, the seismicity occurrence correlation can serve to forecast the general location of the next main earthquake on which fault zone. Furthermore, it offers the preliminary judgment of the pattern of deep evolution of the eastern margin of the Qinghai-Tibetan Plateau, and then from the elementary to the profound, the understanding of characteristics of geodynamic activities.

The eastern margin of the Qinghai-Tibetan Plateau is full of intricate faults systems and a place with many big earthquake events. The characteristics of faults and earthquakes can lead researchers to understand the deep evolution. Densmore *et al.* (2008) used geomorphic observations to analyze the kinematics and slip rates of two active faults that parallel the plateau margin, and concluded that the formation and maintenance of the eastern plateau margin do not involve major upper crustal shortening. P Z Zhang (2013) integratedly studied active faults, GPS crustal deformation, and geophysical structure in the Western Sichuan region, eastern margin of the Tibetan Plateau, and found that the combined model of rigid block movement could be used to describe the continuous deformation of this region. Wang *et al.* (2013) analyzed the crustal P- and S-wave velocity structure and Poisson's ratio recorded along a 1600-km-long profile crossing the southern Tarim basin and then showed a new crustal cross section of the northeastern Tibetan plateau. Most of researches use the priori or inferential theory of tectonic theory and geological characteristics of the regional structure to study the deep construction, evolution and seismology of the eastern margin of the Qinghai-Tibetan Plateau. In our study, we use seismic data to directly study the seismology and conversely use seismology to find out the previously unknown geological characteristics and correlations of the whole faults zones. Then, we conclude a pattern based on these characteristics and correlations and use the pattern to study tectonic theory in this region. This is important because no previous study systematically access the correlations of faults in the eastern margin of the Qinghai-Tibetan Plateau. And since we cannot assert two or more faults with similar strike or dip are belong to the same tectonic stress field, using seismic data to reflect their correlation is justifiable. When calculating the correlations of faults, we have considered the priori combinations, such as Chinese North-South seismic belts or geodynamic blocks in the eastern region, and we use the results of statistical method to compare with this priori combinations. Finally, we get the posteriori pattern of correlations of faults. Knowing these correlations can infer the internal structures and evolutions which cannot be showed by the geological maps.
And practically, compared to the previous methodology of earthquake prediction based on Coulomb stress change, in which the geometrical parameter of faults and total valuables of tectonic stress field are too difficult to be analyzed during the co-seismic process and this uncertainty precludes further credible calculation, the statistical methods analyzing the seismic catalogue can provide a more direct and effective method.
In this study, we use the seismic catalogue as the rudimentary input, bowdlerize the aftershocks using the statistical methods for aftershock exclusion proposed by Uhrhammer (1986), calculate the temporal sequence of Mc, Z and E values in different fault zones respectively as the completeness level, frequency seismicity, and intensity seismicity, in which the Mc value is calculated by Maximum Curvature method (MAXC) (Wiemer and Wyss, 2000) and the Entire Magnitude Range method (EMR) (Woessner and Wiemer, 2005), Z value by Z-test method(R. E Habermann, 1983; R. E. Habermann, 1987) and E value by energy-magnitude relationship (Gutenberg and Richter, 1955). After we get the time-related sequences of the three valuables in different fault zones, we use them to acquire the three correlation coefficients chess table of fault zones though the Pearson's Correlation Coefficient (PCC) and then we quantitatively analyze the correlations based on all of these statistical results.

## 2 Data preprocessing

### 2.1 The complete magnitude of earthquake catalogue

When adopting seismic catalogue data to analyze seismicity, just as this study does, assessing the Magnitude of completeness Mc of seismic catalogue data is an essential and compulsory step, since the choice of Mc is a critical factor (Wiemer and Wyss, 2000). Mc is defined as the lowest magnitude beyond which all earthquakes have been completely and correctly recorded in a space-time volume. After discarding the earthquakes whose magnitude is smaller than Mc, the seismic catalogue data is reliable and suitable for further seismic analysis. Moreover, a correct estimate of Mc is rather fundamental since a small variation in minimum Mc may lead to 25% change of earthquake catalogue volume when b = 1(H Zhang *et al.*, 2016).
Therefore, in this study, the primary task is to properly calculate the minimum magnitude of completeness of earthquake catalogue, by picking appropriate methods for this study. It should avoid underestimation of Mc, which leads to comprising irrelevant even erroneous information and thus to a biased analysis, and meantime should also prevent overestimation of Mc, which brings about under-sampling or even causes the total number of earthquake data insufficient to meet the minimum prerequisite of Gaussian distribution.
Given that using earthquake catalog to assess and quantify the detection capability is complicated and based on many assumptions, which cause uncertainties, the evaluation of the Magnitude of completeness, as the index of detection capability, is challenging. Furthermore, in this study we not only need to analyze the seismicity during a period in a certain fault zone, but also the seismicity correlation between different fault zones. Therefore, considering the temporal distribution of Mc as well as the spatial distribution of Mc is necessary, which make picking the method to obtain Mc even more complex.
There are two distinguished types of methods to calculate Mc: the waveform based method and the statistical method. The waveform-based method, for examples, comparison of amplitude-distance curves, signal-to-noise ratio or amplitude threshold method, defines the probability level by analyzing wave form and its special distribution (D'Alessandro *et al.*, 2011; Ringdal, 1975; Seggern and D., 2004). The most widely applied waveform-based method is the probabilistic magnitude of completeness (PMC) method (Schorlemmer and Woessner, 2008), which introducing detection probabilities with empirical phase-pick data to calculate the interval of possible completeness magnitude. In general, the waveform-based method spends a lot of time on wave form analysis and the introduced probabilities level is only valid for a particular seismic background, so the waveform-based method is not ideal for realistic application.
The statistical method analyzes the relationship between magnitude and frequency. One of the most widely used methods is based on the Gutenberg-Richter (G-R) power law (Gutenberg and Richter, 1944). It assumes that the G-R law distribution is suitable to the occurrence frequency of earthquake magnitudes no less than Mc. Since the G-R law have concrete physics origins and the calculation method is simple, mapping the Mc based on the G-R law is efficient and frequently employed for seismicity analysis (Wiemer and Wyss, 2000; Woessner and Wiemer, 2005).

Mc, as an index of detection capability of seismic stations, varies with time as the technology develops. Therefore, we should select a specific and suitable method to analyze the temporal distribution of Mc and get the average Mc. Moreover, our study region is the area of the eastern margin of the Qinghai-Tibetan Plateau (95°E~107°E, 23°N~36°N),and the seismic stations are spatially unevenly distributed in this area, which leads to the Mc spatially unevenly distributed theoretically.

Considering the spatial variability of Mc is important since in the further analysis, we need to quantify the correlation between fault zones. The premise of reliability of the correlation results is that we give all fault zones equally treatment, which partly means that calculate every fault zone's own Mc. To justify calculating each fault zone's own Mc is necessary instead of redundant, we need to show the spatially uneven distribution of Mc by selecting a specific and suitable method.

When it is applied to realistic data from regional earthquake catalogue, the Entire Magnitude Range (EMR) method is widely used and demonstrates superior performance (Huang *et al.*, 2016). This study applies EMR method to demonstrate the spatial variation of Mc. For spatial distribution of Mc, EMR supposes that the frequency-magnitude distribution conforms to G-R criterion for all events above seismic magnitude of Mc, and conforms to normal distribution for events under seismic magnitude of Mc. So Mc is defined as the critical point where the G-R power law begins to meet (Woessner and Wiemer, 2005).

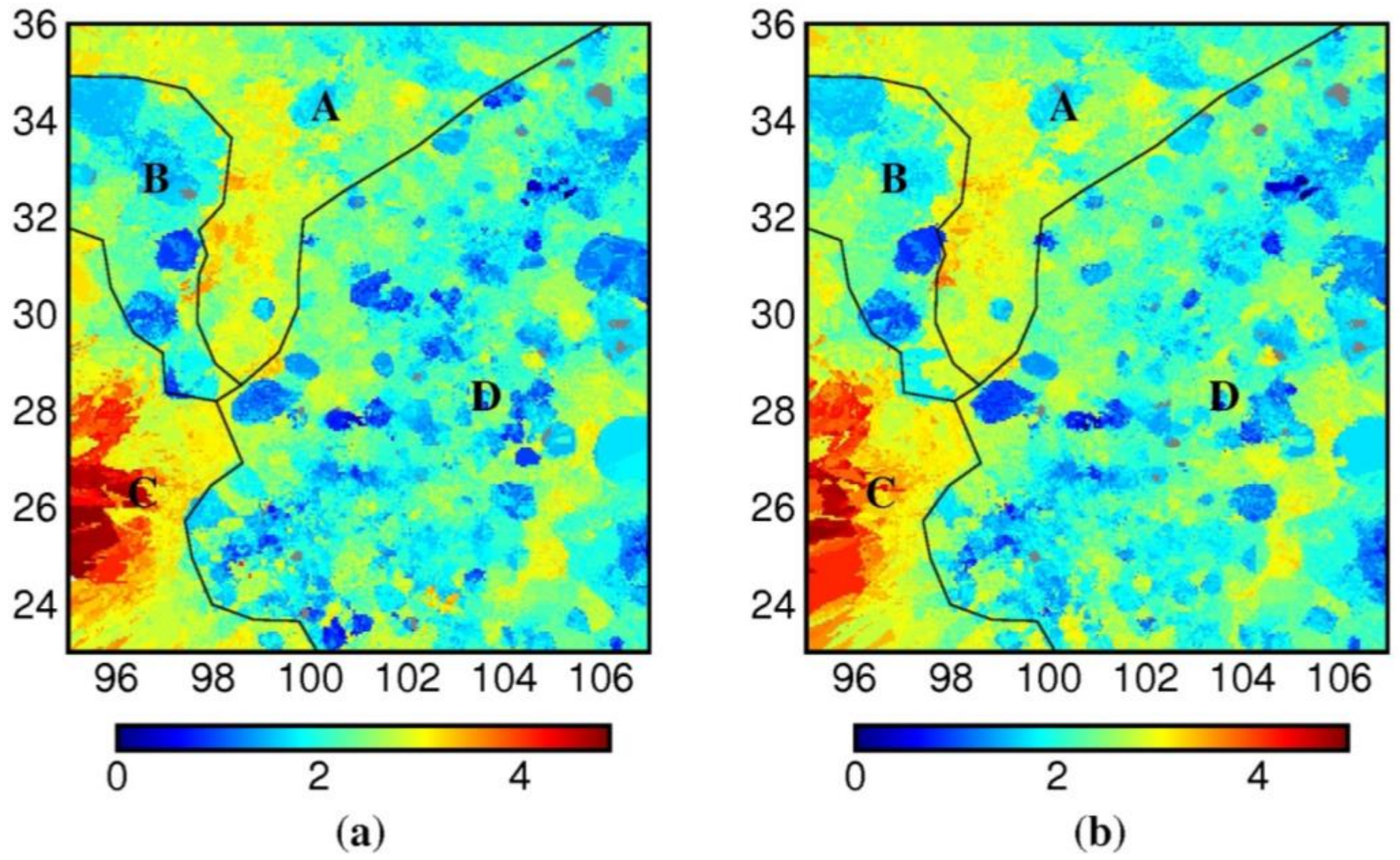

Fig.1 spatial distribution of the complete magnitude of earthquake catalogue calculated by EMR method. (a) Before bowdlerizing the aftershocks; (b) after bowdlerizing the aftershocks.

In the study reg ion (95°E~107°E, 23°N~36°N) and study time period (1970-2015), we gathered 451553 events (from minimal magnitude -1.6 to maximum magnitude 8.0) to calculate Mc values with a 0.05° width uniform square space grid. Fig.1 shows the spatial distribution of the complete magnitude of earthquake catalogue calculated by EMR method before and after bowdlerizing the aftershocks. We can see that even after bowdlerizing the aftershocks, which in a degree discards disturbances, the spatial difference of Mc in our study region is significant. This is because the seismic network is distributed not according to frequency of occurrence of earthquakes, but the development of economy and population. The distribution of seismic network is sparse on the Qinghai-Tibetan Plateau, so the monitoring capacity of earthquakes is limited, which causes the significant bigger Mc value in the western part of our study region. To avoid removing reasonable seismic data in the catalogue, we select four regions A, B, C, D where the Mc value are similar for further processing the catalogue.

As for the spatial distribution, to overcome the disadvantage that the time window truncates some germane earthquake catalogue records, which may cause G-R law to misconstrue the earthquake frequency-magnitude distribution and lead to erroneous Mc value, this study apply the temporal-distribution Mc calculation method proposed by H Zhang *et al.* (2016). They employ a combination of two methods to analyze the temporal distribution of Mc. They are the Maximum Curvature method (MAXC) and the GFT method with 90% (GFT-90%) and 95% (GFT-90%) good-of-fit confidence level. The combination can give a reasonable result of Mc without fit the incomplete part of realistic earthquake catalogue.

Basically, MAXC method calculates the maximum derivative of magnitude-rate curve and chooses the corresponding magnitude as the Mc value. GFT method takes this Mc value as the initial magnitude and repeatedly adjusts the initial magnitude to give a certain evaluation criterion (i.e. in GFT-90%, the evaluation criterion is that a, b and Mc value can explain 90% of the data variability) to estimate the goodness of fit between the observed and synthetic distribution and finally obtains the minimum Mc. And H Zhang *et al.* (2016) also establish a priorities of GFT-95% > GFT-90% >MAXC to determine the final result of Mc.

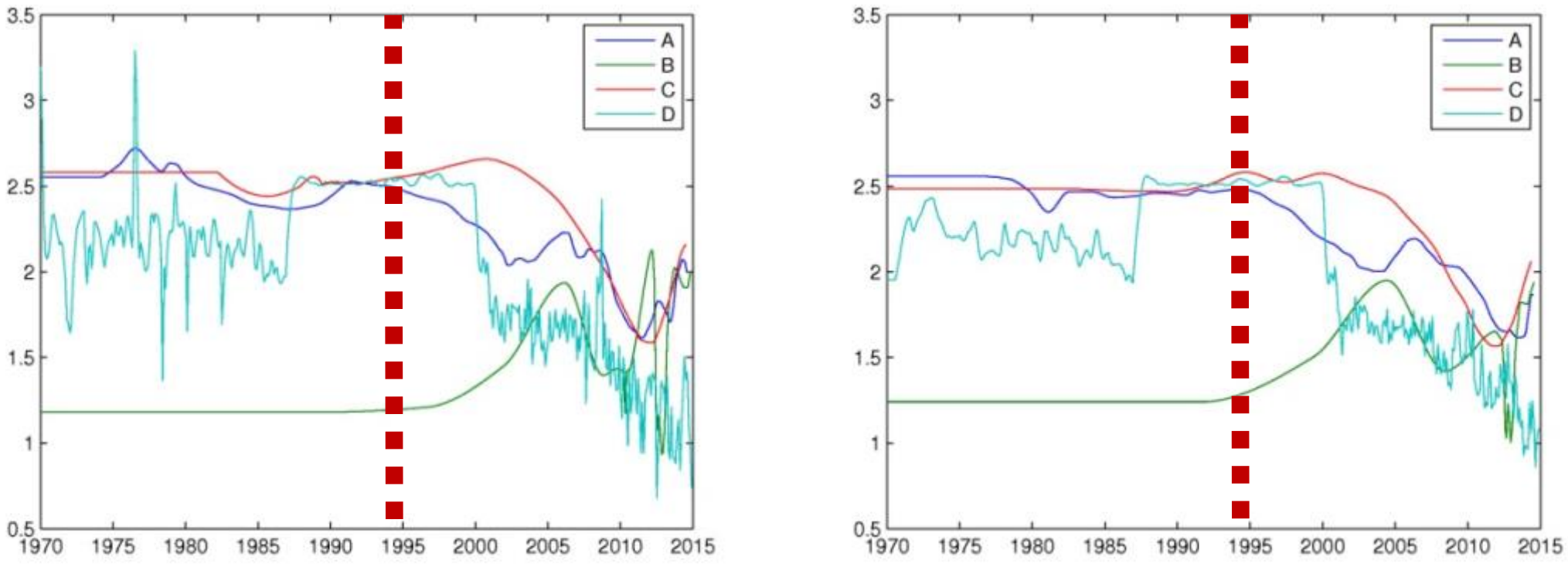

(a) (b)

Fig.2 temporal distribution of the complete magnitude of earthquake catalogue in A, B, C, D regions. (a) Before bowdlerizing the aftershocks; (b) after bowdlerizing the aftershocks.

Fig.2 give the results of the temporal distribution of the complete magnitude of earthquake catalogue in A, B, C, D regions before and after bowdlerizing the aftershocks. Without aftershocks, the temporal changes of Mc get rid of the edges and burrs and become smoother. Around 1995, curves A, B and C begin to decline, which indicate the improvement of seismic network. We can see the smaller and constant value of curve D before 1995, which is caused by the insufficient number of seismic data observed by inadequate local seismic network. While after 1995, curve D starts to ascend to the normal level as other three curves. These changes are consistent with the national project of construction of China Digital Seismic Observatory System and technical transformation of Seismic Precursory Station (Network) during 1995 to 2000.

## 2.2 The complete time of earthquake catalogue

Besides the magnitude of completeness, the time of completeness should also be considered, since insufficient detection ability can miss some earthquakes and the completeness is limited within the period. To measure complete time of earthquake catalogue, the visual cumulative method is proposed by Mulargia and Tinti (1985). This method is based on the assumption that the cumulative number of events increases linearly with time.

In this study, we cumulate the corresponding events whose magnitude are greater than or equal to the Mc value, meaning the catalogue already achieves the completeness with respect to magnitude. Then we get the graph of cumulative number varying with time in the whole study region (Fig.3). In the Fig.3 we can see that after the catalogue being truncated by the Mc value and selected by bowdlerizing the aftershocks, the prepared curve is almost linear. According to Mulargia and Tinti (1985), the time period with an approximate constant slope is consider to achieve the completeness with time.

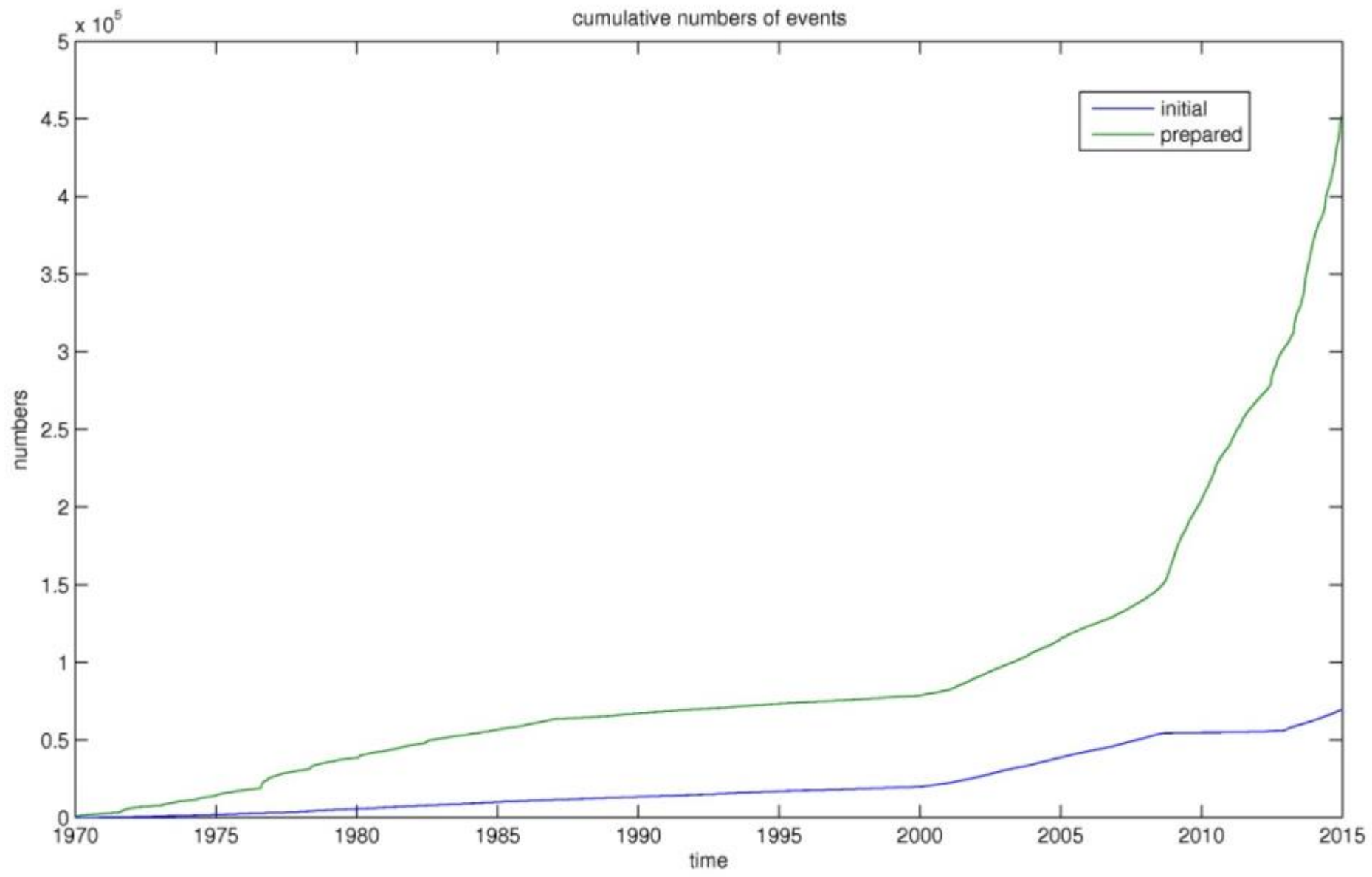

Fig.3 the cumulative number of earthquakes with time. The green line belongs to the original data catalog, while the blue line belongs to the prepared data catalog (being truncated the value below Mc and bowdlerized the aftershocks), which has been truncated by the Mc value and selected by bowdlerizing the aftershocks.

**2.3 The aftershock exclusion**

In this study, we use statistical methods Z-test to measure the seismicity and correlation between fault zones. Since Z-test requires that the data obey a Poisson distribution with a constant Poisson rate, the earthquakes sets calculated by Z-test should be independently occurring. Apparently, aftershocks have strong time-space dependence with main shocks. Hence in order to obtain reliable results, it is indispensable to eliminate aftershocks.

Omori (1984) found the regular temporal distribution of aftershocks denoted as

$$n(t) = \frac{k}{t+c},$$

Where $k$ and $c$ are constants coefficients varying between earthquake sequences. Utsu (1961) refined the empirical relation formula and proposed a modified Omori's law, defined as

$$n(t) = \frac{k}{(t+c)^p},$$

Where $p$ is a constant demonstrating the decay rate.

Based on the modified Omori's law, many statistical methods have been tendered to de-cluster the main shocks and aftershocks and they focus on the choice of spatial and temporal window size. Three major methods are respectively developed by Gardner and Knopoff (1974), P Reasenberg (1985) and Uhrhammer (1986), and all of them assume the time-window and space-window depend on the magnitude of main shock. We chose the de-clustering algorithm developed by Uhrhammer as the most appropriate method because of its simplicity.

Basically, Uhrhammer window method defines the temporal window and space window for all earthquakes based on the modified Omori's law and their magnitude, and finds clusters of the dependent events fall within the windows, and finally removes the dependent events and regards the epicenters of these clusters as main shocks.

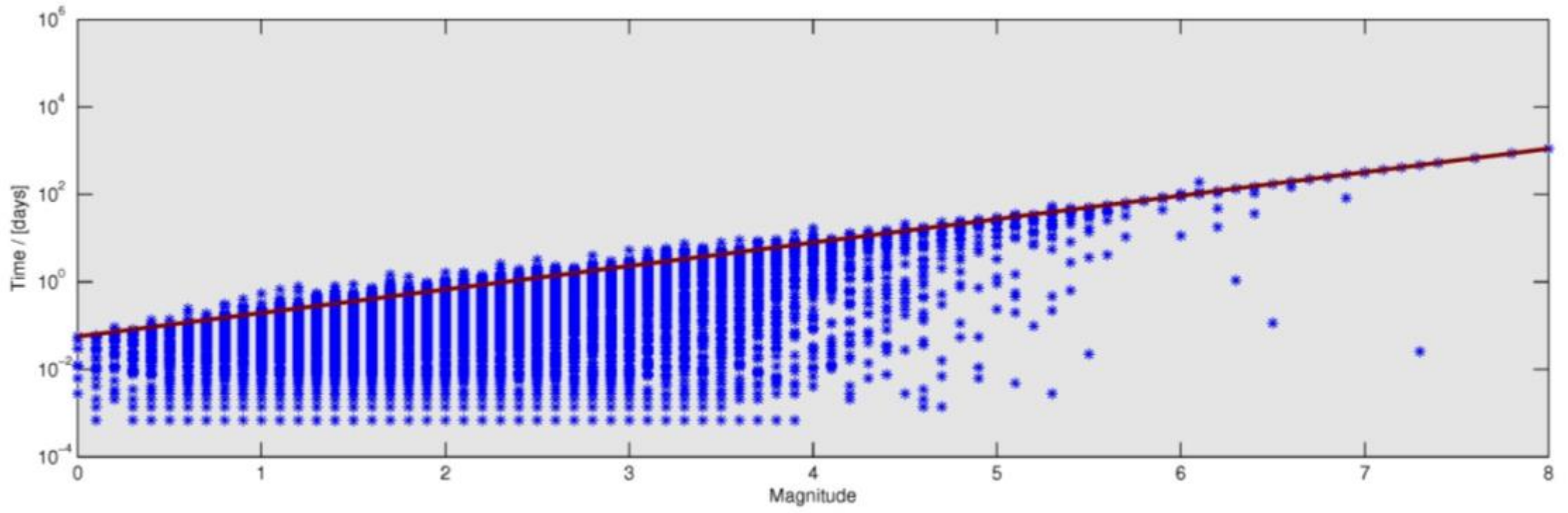

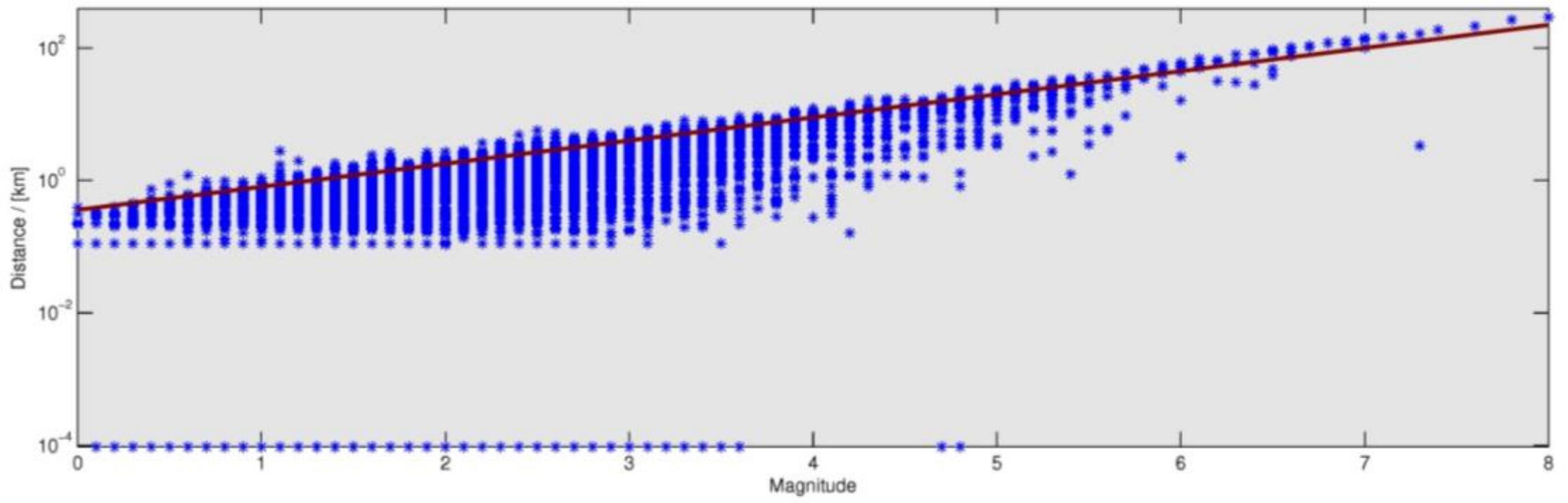

Fig.4 Graph showing the magnitude-dependent space and time windows used for the de-clustering of foreshocks and aftershocks (dependent events) using the method of Uhrhammer (1986). The blue dots that fall below the window lines are considered to be dependent events.

## 3 Methodology

### 3.1 Z-test method

The two special states of the seismicity are seismic quiescence and seismic activation. To quantify the significance of difference of these two states relative to the background seismicity, the concepts of seismic frequency and seismic intensity change are introduced. When it comes to comparing the seismic frequency between two time periods, seismicity rate change is a suitable parameter to demonstrate the seismicity increase or decrease.

Theβ-statistic method (Matthews and Reasenberg, 1988; P A Reasenberg and Simpson, 1992) gives the value of β, by dividing the difference of number of events following an event and the expected number (assuming seismicity is stationary) by the variance of the expected number. The large and positive value of βto stand for the seismicity increase. But the difficulty in assessing the statistical significance of any β-statistic result makes theβ-statistic method untoward. Extending the β-statistic method, David Marsan and Nalbant (2005) develop P-statistic method and γ-statistic method (D. Marsan, 2003) to solve the intractable problem on the statistical significance. It defines that |γ|=1.6 responds to 95% confidence level, |γ|=2 responds to 99% confidence level.

Another statistic method to measure both the seismicity rate change and the significance of the change is Z-test statistic method proposed by R. E Habermann (1983). Z-test statistic is based on the classic statistical method named Z-test, which test whether two independent populations have the same mean. When Z-test is applied in seismicity rate change, it tests the significance of difference between the frequency Poisson distributions of two independent earthquake samples from the same sequences. It defines that |Z|=1.64 responds to a 90% confidence level, |Z|=1.96 responds to 95% confidence level, |Z|=2.57 responds to 99% confidence level (R. E. Habermann, 1987).

Under the usages of many previous studies, the Z-test method show a capability of efficiently assemble the complicate earthquake samples from the regional seismicity sequence both in time windows and space windows, and Z-test based classical statistical method strictly measures the significance (Toda *et al.*, 1998; Wiemer and Wyss, 2000; H Zhang *et al.*, 2016) .

Therefore, in this study, we adopt the Z-test to measure the seismicity rate change and the significance of the change. Before calculating the Z value, we need to evenly discrete the two intervals $[T_1, T_2]$ and $[T_1 + \theta_1, T_2 + \theta_2]$ of the two earthquake samples by the time bin length $\theta$. The numbers of bins respectively are $m_1 = (T_2 - T_1)/\theta$ and $m_2 = \left((T_2 + \theta_2) - (T_1 + \theta_1)\right)/\theta$, the sequence of their number of earthquakes in corresponding bins are $\left\{n_1^{(1)}, n_1^{(2)}, \cdots, n_1^{(m_1)}\right\}$ and $\left\{n_2^{(1)}, n_2^{(2)}, \cdots, n_2^{(m_1)}\right\}$ with the mean $\mu_1$and $\mu_2$ and standard deviations $\sigma_1$ and $\sigma_2$. Then the Z value is defined as

$$Z = \frac{\mu_1 - \mu_2}{\sqrt{\frac{{\sigma_1}^2}{n_1} + \frac{{\sigma_2}^2}{n_2}}}$$

With $n_1$and $n_2$ being the total number of two earthquake samples. Note that $n_1$and $n_2$ are both Poisson random variables normally distributed by central limit theorem. The premise of Z-test method requires that the data sets is from a stationary Poisson process. In this study, the number of earthquakes in eastern margin of the Qinghai-Tibetan Plateau is large enough to meet the requirement of being Poisson-distributed.

### 3.2 Correlation coefficients calculation

In this study, the correlation coefficients calculation method is the widely used Pearson's Correlation Coefficient (PCC), which is a product-moment using the ratio of two data's covariance and the product of their own standard deviation. And we will set the 95% confidence level and calculate P-value to assess the initial hypothesis that the two sequences are not relevant.

To consider the factors causing the differences between fault zones as completely as possible, we introduce two time-dimension parameters to calculate the correlation coefficient Z value and E value, representing seismic frequency and seismic intensity respectively. And we choose 95% significance level to assess the significant and non-significant correlations.

As for the correlation coefficient of Z value, its purpose is to accurately assess the two fault zones' similarity in the seismicity change. Since the seismicity change has the well-known periodic recurrence characteristic, it is important to decide the time interval sequence. Many previous work on Z value choose the same time interval before and after an event to estimate the seismicity change (Jafari, 2012; Sorbi *et al.*, 2012), and this is suitable to assess only one event considering the need of variable control approach. But to achieve a statistically significant correlation coefficient, it is necessary to assess a lot of events and obtain enough amount of Z value in a long time containing many recurrence periods in different fault zones, which makes choosing time intervals complicated. Because considering the periodicity of seismicity change makes the correlation coefficient contain the important information of the similar or different responses to same events (big earthquakes) in different fault zones, which makes the completeness of earthquake periods weigh more than strict evenness of time interval. Besides, Z-test allows the length of time intervals to be different, as long as the time bin length is the same.

In this study, a period of seismicity change is the time interval between two events whose magnitude is bigger than or equal to 6.0, and the time bin length is 14 days. Then we get a time-ordered time interval sequence divided by Mw≥6.0 events, and we will calculate the Z value of two adjacent time interval. Moreover, to get rid of the interference of the variability of different fault zones' size, we set a uniform square space grid with 0.05°width and get Z-value in every cell, and then calculate the weighted average of Z values in all cells as the final Z value in $\mathrm{i^{th}}$ pair of two adjacent time interval. The time-dimension sequence of the weighted average of Z values is denoted by $\left\{Z_{x_1}, Z_{x_2}, \ldots, Z_{x_i}, \ldots, Z_{x_n}\right\} (i = 1, 2, \ldots, n)$, where n mean the $\mathrm{n^{th}}$ pair of two time adjacent intervals.

We use all the weighted average of Z values to assess the correlation of different fault zones. The correlation coefficient of Z value is defined as

$$R_Z = \frac{\sum_i \left(Z_{x_i} - Z_{x_m}\right)\left(Z_{y_i} - Z_{y_m}\right)}{\sqrt{\sum_i \left(Z_{x_i} - Z_{x_m}\right)^2} \sqrt{\sum_i \left(Z_{y_i} - Z_{y_m}\right)^2}}$$

where $Z_{x_i}$ and $Z_{y_i}$ are the respective $\mathrm{i^{th}}$ element in the sequence of weighted average of Z values in the fault zone X and fault zone Y, $Z_{x_m}$ and $Z_{y_m}$ are corresponding arithmetic mean value of the two sequences. The closer $|R_Z|$ is to 1, the more related the two fault zones are.

In the similar way, the calculation of correlation coefficient of E value uses the same time interval sequence and space grid, while it gets a E value within only a time interval, which means the sequence of E values has one more element than the sequence of Z values. The time-dimension sequence of E values in fault zone X is denoted by $\left\{E_{x_1}, E_{x_2}, \ldots, E_{x_i}, \ldots, E_{x_n}, E_{x_{n+1}}\right\}(i=1,2,\ldots,n,n+1)$. In each time interval, the E value in a fault zone is defined as the ratio of cumulative energy of all the earthquakes and the number of the cells in the fault zone. The energy of a single earthquake (Gutenberg and Richter, 1955) is defined as

$$\log e_j = 11.8 + 1.5 M_j$$

where j means the j$^{th}$ earthquake within a time interval. And in i$^{th}$ time interval, the E value is defined as

$$E_i = \frac{\sum_j e_j}{N}$$

where N means the amount of the space grid cells within a fault zone. Then, we calculate the change of $\Delta \log_i(E)$ between two adjacent time intervals:

$$\Delta \log_i(E) = \log(E_{i+1}) - \log(E_i)$$

which represents the change of magnitude of energy. Then we can get a sequence $\left\{\Delta\log_{x_1}(E), \Delta\log_{x_2}(E), \ldots, \Delta\log_{x_i}(E), \ldots, \Delta\log_{x_n}(E)\right\}(i=1,2,\ldots,n)$. And the correlation coefficient of E value is defined as

$$R_E = \frac{\sum_i \left[\Delta\log_{x_i}(E) - \Delta\log_{x_m}(E)\right]\left[\Delta\log_{y_i}(E) - \Delta\log_{y_m}(E)\right]}{\sqrt{\sum_i \left[\Delta\log_{x_i}(E) - \Delta\log_{x_m}(E)\right]^2}\sqrt{\sum_i \left[\Delta\log_{y_i}(E) - \Delta\log_{y_m}(E)\right]^2}}$$

where $E_{x_i}$ and $E_{y_i}$ are the respective i$^{th}$ element in the sequence of E values in the fault zone X and fault zone Y, $E_{x_m}$ and $E_{y_m}$ are corresponding arithmetic mean value of the two sequences. The closer $|R_E|$ is to 1, the more related the two fault zones are.

Moreover, the major factor leads to our results deviating from the real geological correlations of fault zones is the systematic error, which could be the uneven spatial distribution or the variable detection capability of seismic network.

A small earthquake is less likely to be detected where the seismic network is distributed sparsely, which may lead to only parts of events are recorded even when the same-magnitude earthquake actually occurs in different fault zones. This can make the correlation coefficient of Z value $R_Z$ cannot fully show the correlation of different fault zones with different density of seismic network distribution.

And the improvement of detection capability makes the detected magnitudes more accurate, but the building of the improved seismic networks is spatially variable and doesn't get rid of the old stations, which makes the different fault zones record different values of magnitudes even in fact they are same. This can make the correlation coefficient of E value $R_E$ cannot fully show the correlation of different fault zones with different quality of seismic network.

Therefore, to minimize these systematic errors, we introduce the correlation coefficient of Mc value with time to assess the deviation and assist with the analysis of geological correlation of fault zones. In the similar way as $R_Z$ and $R_E$, the calculation of correlation coefficient of Mc value uses the same time interval sequence and space grid. The method to obtain the Mc-time curves of different fault zones is MAXC method with GFT-90% and GFT-90%, which has been introduced in section 2.1. The time-dimension sequence of Mc values in fault zone X is denoted by $\left\{Mc_{x_1}, Mc_{x_2}, \ldots, Mc_{x_i}, \ldots, Mc_{x_n},\ Mc_{x_{n+1}}\right\}(i=1,2,\ldots,n,n+1)$. The correlation coefficient of Mc value is defined as

$$R_{Mc}=\frac{\sum_i\left(Mc_{x_i}-Mc_{x_m}\right)\left(Mc_{y_i}-Mc_{y_m}\right)}{\sqrt{\sum_i\left(Mc_{x_i}-Mc_{x_m}\right)^2}\sqrt{\sum_i\left(Mc_{y_i}-Mc_{y_m}\right)^2}}$$

where $Mc_{x_i}$ and $Mc_{y_i}$ are the respective $\mathrm{i}^{\mathrm{th}}$ element in the sequence of weighted average of Mc values in the fault zone X and fault zone Y, $Mc_{x_m}$ and $Mc_{y_m}$ are corresponding arithmetic mean value of the two sequences. The closer $|R_{Mc}|$ is to 1, the smaller the systematic errors is caused. We use the combination of the correlation coefficients $R_Z$ and $R_E$ to preliminarily estimate the seismically correlation in the aspects of both the seismic frequency and the seismic intensity between different fault zones. Then, we value the reliability of the seismically correlation with the correlation coefficient $R_{Mc}$, with which is close to 1 the result of the seismical correlation is less disturbed by the systematic error and is more reliable.

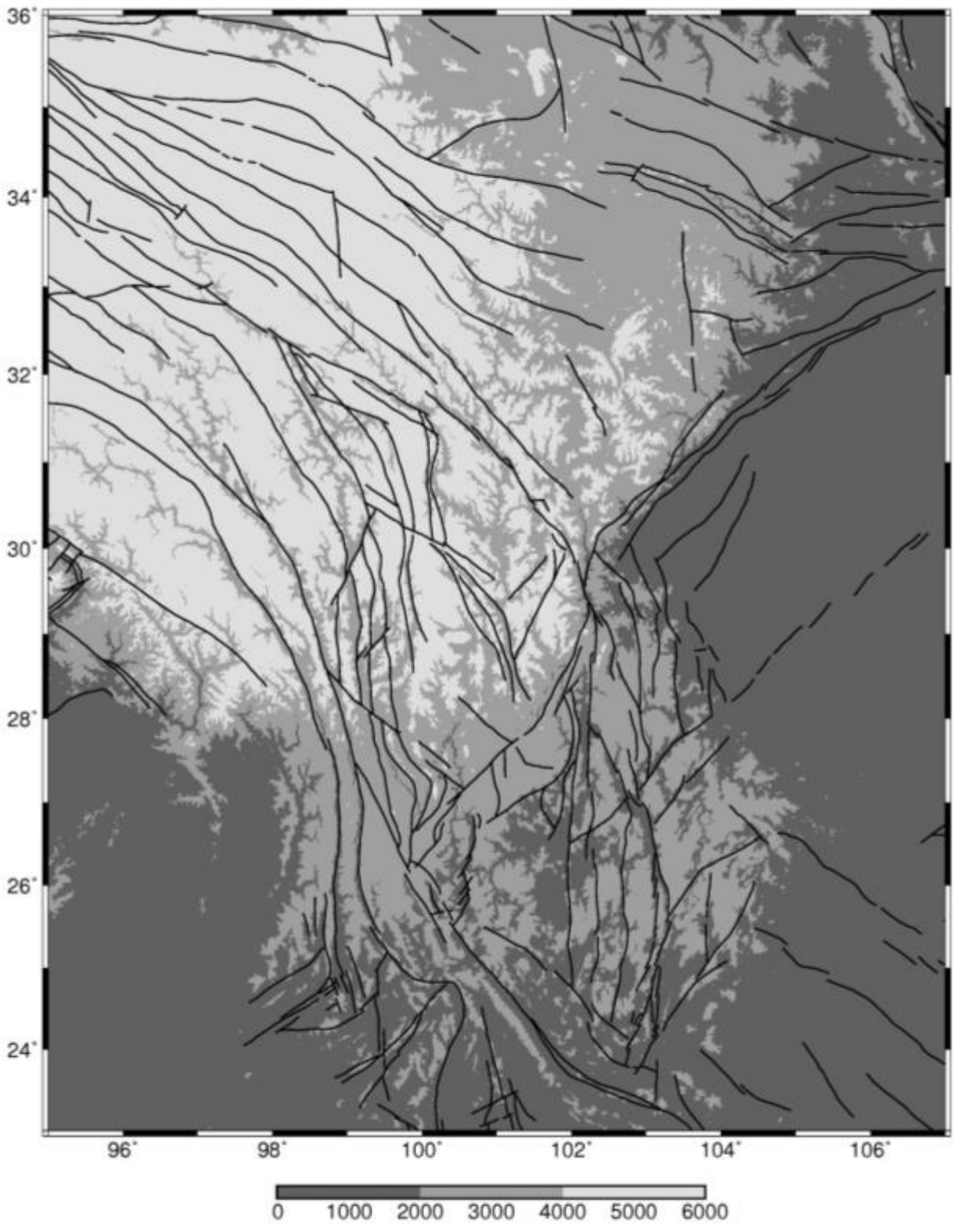

Fig.5 the location of research region.

**4 Regional geologic setting and divisions in the eastern margin of the Qinghai-Tibetan Plateau**

Generated by the India-Asia collision since the late Quaternary period, the Qinghai-Tibetan Plateau has become the highest and biggest tectonic and physiognomy union in the whole world. And the eastern margin of the Qinghai-Tibetan Plateau has the most complicated tectonic activities and the highest density of active faults. This eastern margin usually means the area between eastern Himalaya belt and west Qinling orogenic belt, surrounded by the East Kunlun fault system, Longmenshan tectonic belt, and Lancang River Fault zone respectively as its northern, eastern and southwestern boundary.

The structural deformation pattern, dynamic mechanism and its relations to big earthquake events in the eastern margin of the Qinghai-Tibetan Plateau have received great concern. In respect to the warning of disaster, these attributes can be used to assess seismic or landslide hazard, which is conducive to a safer residential environment and a more secure infrastructure; In respect to the function of geophysics research, these attributes are believed to be the key to solve the genetic mechanism of the Qinghai-Tibetan Plateau even the whole East Asia continent.

The structure in the eastern margin of the Qinghai-Tibetan Plateau is unique and instructive. Since the stiff rock of Sichuan basin stymied the eastward entrench of Qinghai-Tibetan Plateau, the amassing energies are displayed in the way of faults and earthquakes. Practically, there are five important seismic belts from the well-known Chinese North-South seismic belts in this area: the Xianshuihe-Anninghe-Xiaojiang seismic belt, Jinshajiang-honghe seismic belt, Lancangjiang seismic belt, Yaluzangbujiang-Nujiang seismic belt and Longmenshan seismic belt.

Seismic is active and concentrated in the seismic belts and the combination of them lies north and south down the Chinese southern border. While all the shapes of five seismic belts are same actuate, they have varied fault zone numbers, different seismicity, diverse fault attributes, due to the complicated genetic mechanism of the eastern margin of the Qinghai-Tibetan Plateau. Hence, it is necessary to dissect and classify the fault zones.

In this study, we focus on all the main faults in the eastern margin of the Qinghai-Tibetan Plateau especially the ones in the five important seismic belts. Since the scopes of the five seismic belts are well overlapped with the corresponding scopes of five fault belts, we use fault belt to refer to seismic belt.

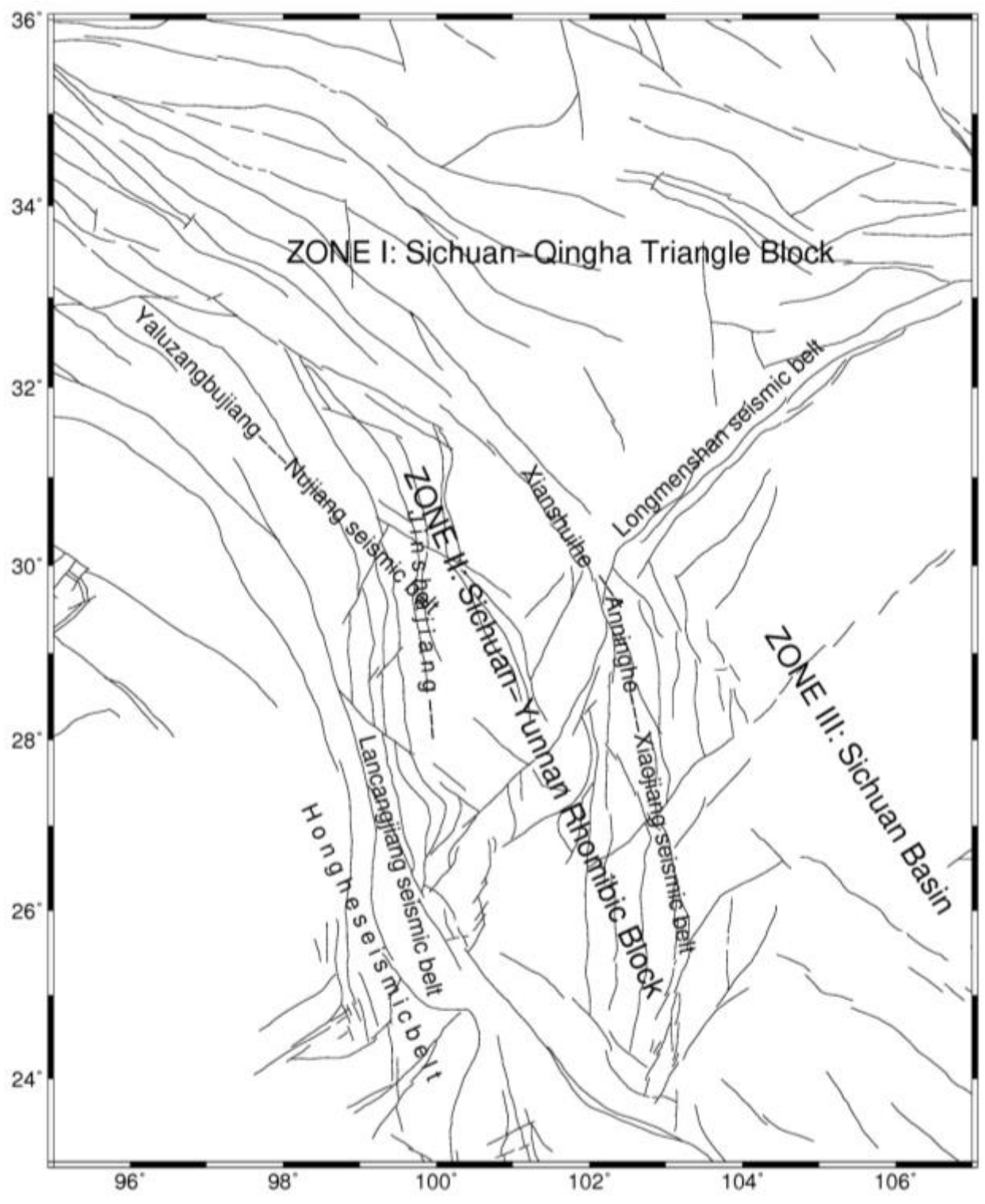

Fig.6 The main seismic belts and geodynamic blocks

As the Fig.6 shows, in the center of the study region, two important fault belts, Xianshuihe-Anninghe-Xiaojiang fault belt and Jinshajiang-honghe fault belt, form the special structure Sichuan-Yunnan rhombic block. Along the boundary of the rhombic block all the active faults belonging to the two fault belts display a feature of high speed left-lateral slipping, leading to the clockwise rotation of the rhombic block; within the rhombic block lies two fault zones, Litang fault zone and Muli fault zone, which are also the left-lateral slipping faults, splitting into three clockwise sub-blocks inside of the block. Because of the particularity of coincident left-lateral character, the rotary block mode (Zhou *et al.*, 2006) is proposed to demonstrate the main dynamic mechanism of the eastern margin of the Qinghai-Tibetan Plateau.

In fact, in this region, there is another block called the Sichuan-Qinghai block, which is a triangle block adjacent to the Sichuan-Yunnan rhombic block taking the Xianshuihe fault zone as the separatrix. The Sichuan-Qinghai block is confined by the Yushu fault zone and Xianshuihe fault zone as its west-southern edge, Longmenshan fault belt and Longriba fault zone as its east-southern edge, Aba fault zone and Dongkunlun fault zone as its northern edge. It's worth noticing that these fault zone on the edges are all left-lateral except for the longmenshan and Longriba fault zones. What makes these two fault zones special is that he Longmenshan fault zone together with the Minjiang-huya fault zone bears the south-eastward deformation from the Qinghai-Tibetan Plateau uplift, and the Longriba fault zone may resolve this deformation by the form of thrust left-lateral slip faults at the Longmenshan fault zone's trailing edge.

In the south-eastern part of the study region lies the Sichuan basin, where the terrain and the number of faults are acutely declined. The only main fault zone is the Huarongshan fault zone, which is almost parallel with the Longmenshan fault zone and thrust right-lateral slip faults as the same. It can presume that the Huarongshan and Longmenshan fault zone may have the same dynamic mechanism though in different geological backgrounds.

The eastern tail of Yaluzangbujiang-Nujiang fault belt is in the western part of the study region, and the tail is a combination of Nujiang fault zone, Lancangjiang fault zone, Dayinjiang-ruili fault zone, Longlin fault zone. From the Fig.6, the geometrical shape of the tail well fit the western edge of the Sichuan-Yunnan rhombic block. Hence, it is highly possible that the genetic mechanisms of the Yaluzangbujiang-Nujiang fault belt and the Sichuan-Yunnan block are closely related.

To justify the inherent rotary block mode in the Sichuan-Qinghai triangle block, Sichuan-Yunnan rhombic block, or to clarify the similarity of dynamic mechanisms of Longmenshan fault zone and Huarongshan fault zone, or to figure out the how the genetic mechanisms of Yaluzangbujiang-Nujiang fault belt and Sichuan-Yunnan block influence each other, it is positively necessary to analyze the correlation of all the fault zones.

In this study, we use the seismic catalogue to measure the main fault zone's characteristic and their correlation. We select a space window of 23°N-36°N, 96°E to 105°E, and a time window of year 1970 to year 2015 to calculate the seismicity of 21 faults in the study region (as the Fig.7 showed). Besides, we use the correlation method (proposed in the section 2.3) to measure the correlation of the 21 faults, and compare our result with the combination of the five important fault belts.

The eastern margin of the Qinghai-Tibetan Plateau (23-36N; 95-107E) is a desirable area for our study region for the following reasons:

1. In the aspect of its geotectonic setting, the eastern margin of the Qinghai-Tibetan Plateau is relatively spatially closed, located in the long narrow "bottleneck" part of the Gansu-Qinghai-Tibet plate. It is west and south adjacent to the India plate and east conterminous to the Southern China plate, which justifies the area to ignore the effects of fracture movements outside the region and only takes into account the influence from internal earthquakes.
2. The eastern margin of the Qinghai-Tibetan Plateau is a very seismically active region in the Eurasian plate, influenced by the collision from the Indian plate, and its earthquakes occur frequently and faults are well developed.
3. The eastern margin of the Qinghai-Tibetan Plateau has few other disaster events, such as outbreak of volcano or massive landslides, can cause earthquakes. So, the seismic catalogue can exclusively be attributed to the movements of fault zones, which ensures the reliability of using the earthquake catalogue to analyze the correlation of fault zones.
4. So far, no seismic risk assessment analysis considering the correlation of fault zones has been performed for the eastern margin of the Qinghai-Tibetan Plateau. Our study definitely can provide new perspectives and angles to this area.

The fault zone division follows the descriptions from three authoritative works of Cheng *et al.* (2011); Shen *et al.* (2003); Shen *et al.* (2005), Our principle for fault zone distribution is

1. The fault zone includes the main fault and the faults parallel to it.
2. The division borders try to parallel to or perpendicular to the main fault.
3. The adjacent fault zones use the same border, meaning there is no gap between fault zones.

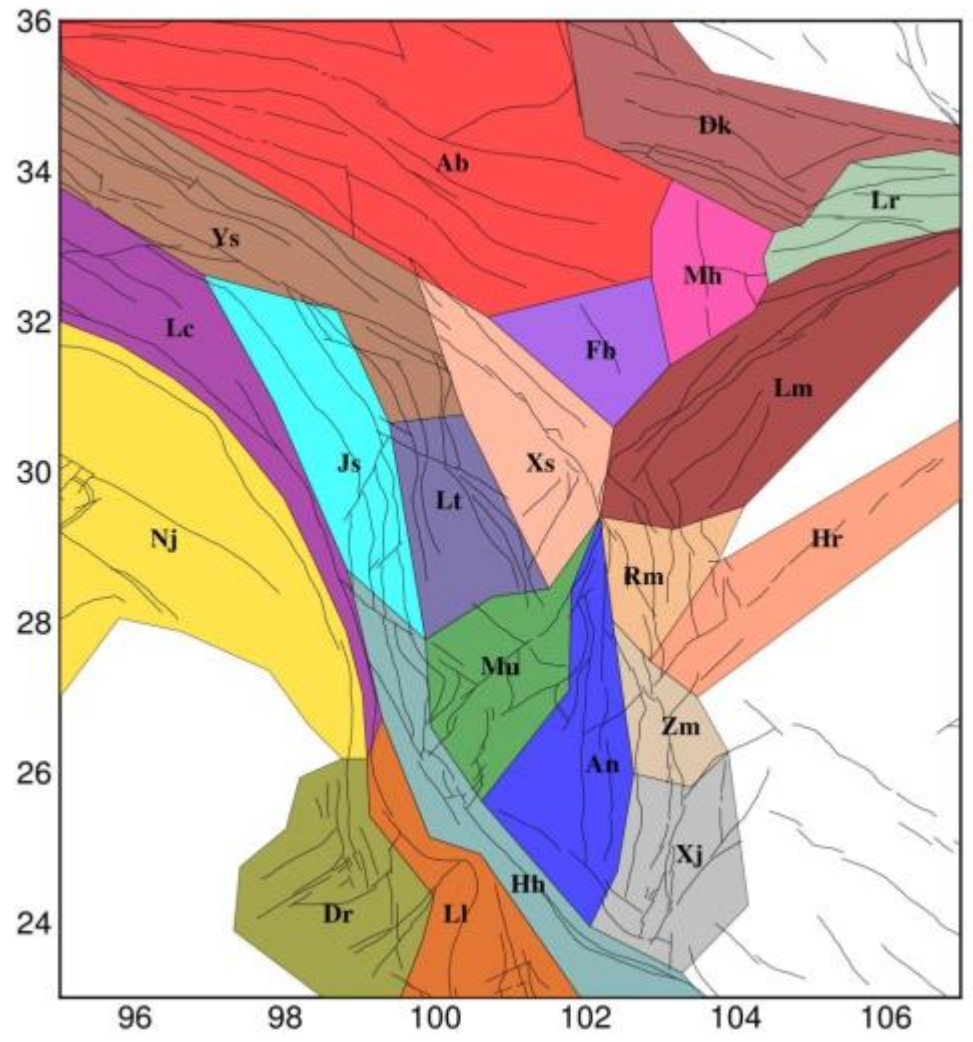

Fig.7 The distribution of 21 main faults zones in the eastern margin of the Qinghai-Tibetan Plateau. Ab:Aba fault zone; An: Anninghe fault zone; Dk: Dongkunlun fault zone; Dr: Dayingjiang-ruili fault zone; Fb: Fubianhe fault zone; Hh: Honghe fault zone; Hr: Huarongshan fault zone; Js: Jinshajiang fault zone; Lc: Lancangjiang fault zone; Ll: Longlin fault zone; Lm: Longmenshan fault zone; Lr: Longriba fault zone; Lt: Litang fault zone; Mh: Minjiang-huya fault zone; Mu: Muli fault zone; Nj: Nujiang fault zone; Rm: Rongjing-mabian fault zone; Xj: Xiaojiang fault zone; Xs: Xianshuihe fault zone; Ys: Yushu fault zone; Zm: Zemuhe fault zone.

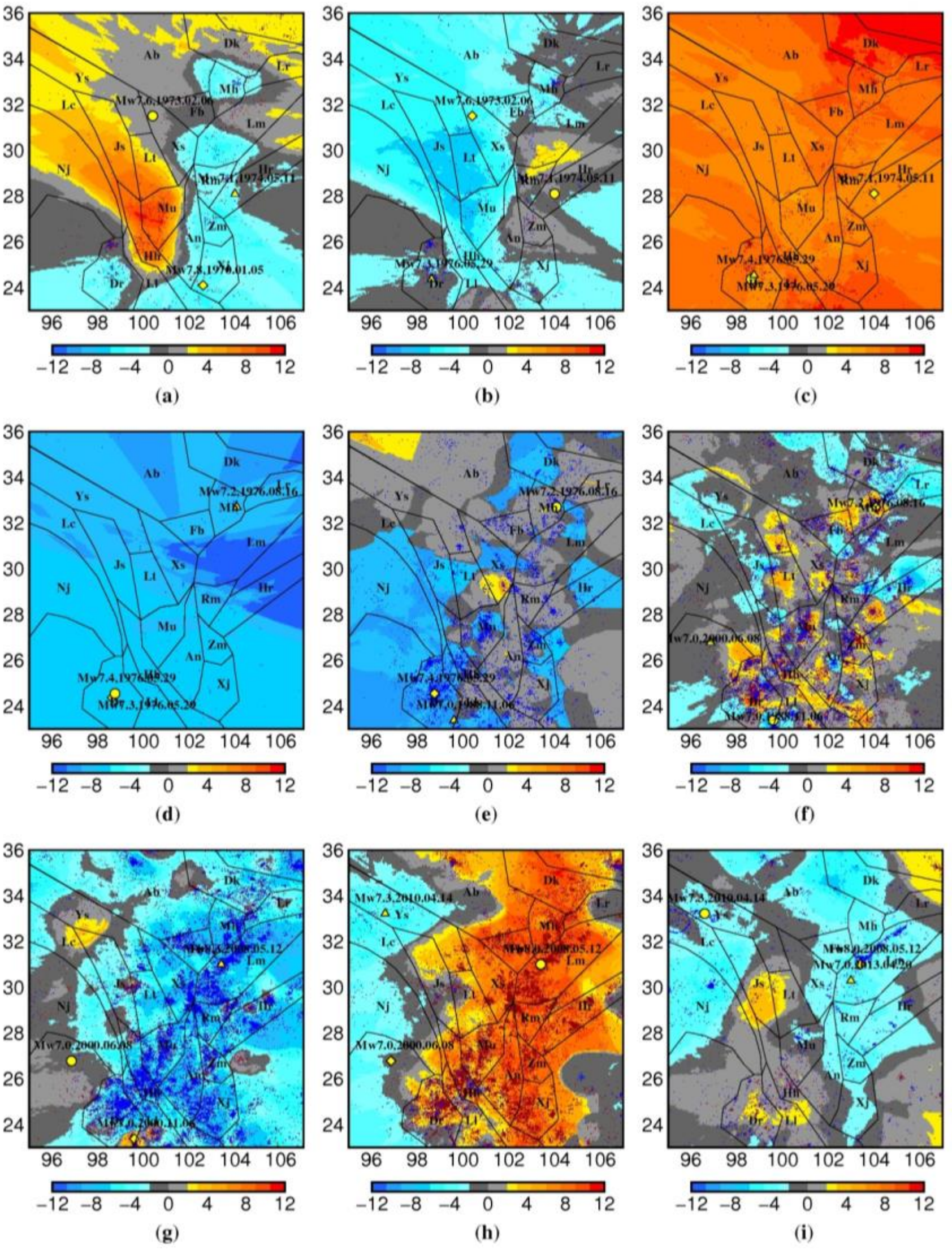

Fig.8 The spatial distribution of seismicity rate change in 9 pairs time sequence from adjacent 11 time intervals, using z-test method. The 11 time intervals are divided by Mw > 7.0 earthquake events. In each sub-map, three big events involve, the previous, middle and latter events, which are respectively marked by diamond, circle and triangle symbols. The red points show the locations of earthquakes dated from the previous events to the middle events, while the blue points shows the locations of earthquakes dated from the middle events to the latter events. The negative Z-value means seismicity increase, denoted by blue color; while the positive Z value means seismicity decrease, denoted by warm color. We choose 95% significance levels and paint gray when the area owns a |Z| value lower than 1.96.

## 5 Results

### 5.1 Results of Z-test method

We have selected all eleven Mw>7.0 events from the earthquake catalogue in eastern region of Qinghai-Tibetan Plateau (23-36 ° N, 95-107 ° E) from 1970 to 2015, and use each two adjacent occurrence time intervals of the eleven Mw>7.0 events to confine an interval, so we get 10 intervals and 9 pairs of adjacent intervals to address seismicity rate changes by using Z-test method.

The time bin is 14 days by convention, and the space grid is 0.05 ° which is consistent with all the calculations in this study. And we use the no less than 100 events within a radius of 50 km for each node to calculate Z value. As the section 3.1 proposed, a positive Z value denotes seismicity decrease, while a negative Z value represents seismicity decrease.
We obtain 9 Z value maps (Fig. 8), in each sub-map we use red points to show the earthquakes events in the previous time interval and blue points the latter time interval, and use different symbols to distinguish the previous, middle, posterior events, respectively represented by diamond, circle and triangle. The 9 Z value maps follow the order of time and imply both the change between two big events and the cumulative change. We also mark the location of the three ends (the occurrence time of Mw>7.0 big events) of the two adjacent time intervals. Since Z value has self-supplied criteria to access the significance levels, namely |Z| values of 1.64, 1.96 and 2.57 corresponding to 90%, 95% and 99% significance levels. In this study, we choose 95% significance levels and paint gray when the area owns a |Z| value lower than 1.96. We regard the dark gray and light gray area as a stationary state in the area.
In the Fig.8(a), since the magnitudes of the previous event (diamond) and the middle event (circle) is similar (Mw 7.8 and Mw 7.6), so we can consider the influence of distance. In region Js, that large area is full of warm color demonstrates that most area in Jinshajiang fault zone become inactive in seismicity after the middle event, which is near the region Js. The same condition also happens in region Lc, region Lt, the north-western part of region Ys and the northern part of region Nj. This condition is equivalent to the condition that the seismicity increases (the color is blue) while the area is much closer to the previous event than to the middle event, such as in the region Xj, region Zm, region Dr, eastern part of region An and northern part of region Ll.
This shared characteristic could conclude a basic pattern: close distance to the epicenter leads to the seismicity decrease.
This energy theory could explain the pattern. A big event rapidly releases a great deal of stored potential energy, and it needs more time to accumulate enough energy to trigger other earthquakes, so in the area around the big event the following small events are relatively rare.
But the basic pattern is not suit for every fault zone. In region Ab, that large area is light gray and small area is yellow demonstrates that most area in Aba fault zone keeps an unchanged seismicity and the west-northern area become slightly inactive in seismicity. Although the middle event is much closer to the Aba fault zone than the previous event, the Aba fault zone isn't greatly affected by the middle event. The same condition also happens in region Fb, the north-western part of region Xs and the south-eastern part of region Ys. Moreover, the seismicity change of region Mh is the extreme case contradicting the pattern-1. Region Mh is full of blue indicating seismicity increase while it is close to the epicenter of the middle event.
Their specialty may suggest there is another factor that influences the seismicity. In this paper, we assume the factor is the different correlations between each fault. The seismicity changes in region Mu and region Rm can support our assumption. Even though both of the two regions have almost equal distances to the previous event and middle event, these two regions display a significant seismicity change: seismicity decreases in region Mu while increases in region Rm.
Furthermore, the close distance and sharp difference in seismicity change between region Mu and region Rm also indicate that in some area the internal intimacy weigh more than external distance.
In Fig.8(b), following the basic pattern and considering the potential factor of internal correlation, we can find the commonness as in Fig.8(a). The area around the middle event shows seismicity unchanged or decreasing, such as region Hr, region Rm, region Zm, region Lm; while the region surrounding the previous event performs seismicity increase, such as region Mh, region Fb, region Xs, region Lt, region Js, region Lc, region Ys, region Ab. Generally, Fig.8(b) complies the basic pattern, but while region Mu, region Hh, region Ll and region Nj are all far away the previous and middle event, they are all blue, meaning significant seismicity increase.
Since we regard the big event (Mw > 7.0) as a disturb to a stable seismic background field, the far area, such as region Dk and region Lr and region Dr, shows a gray color meaning stationary seismicity is normal. However, region Mu, Hh, Ll and Nj are equal far away to or even more far away than region Dk and Lr, but they still show a blue color meaning seismicity increase. These abnormities indicate some other internal mechanism between certain fault regions have more influence on the seismicity than the distance.
Fig.8(c) and Fig.8(d) is conspicuously different from others, showing significant seismicity change in whole area because of the sudden temporal and spatial concentration of two big event. The results of Z test intuitively demonstrate seism's anomalous behavior in the three intervals, involved four big events: the first Mw 7.1 big earthquake event in 1974.05.11, second Mw 7.3 in 1976.05.29, third Mw 7.4 in 1976.05.29 and fourth Mw 7.2 in 1976.08.16. Because the occurrence time of second big

event and third big event are highly close, the cumulative number of main shocks is relatively small, resulting in the relatively low seismicity compared to the two other intervals. The binary events happened on 1976.05.29 at Longlin County in Yunnan Province are characteristic by their almost simultaneous occurrence, alike magnitude and contiguous location. The formerly happened Mw 7.3 main shock comes from a NW trending dextral reverse fault, the later happened Mw 7.4 main shock comes from a NE trending sinistral reverse fault, and both of the main shocks happened in the Dayingjiang-Ruili fault zone, which belongs to a triangle block confined by Yaluzangbujiang-Nujiang seismic belt and Honghe seismic belt. From the Fig.8(d) to Fig.8(h), we can see that after the two special earthquakes, the Dayingjiang-Ruili fault zone display a cumulative seismicity increase until the end of the super Mw 8.0 earthquake in 2008.05.12. The abnormity of the sudden large-area change of Z value may indicate the concentration of two big events, which leads to seismicity keeping a continuous long-term active state.

Fig.8(e) generally shows a relatively stable seismicity in large area, indicating the middle event didn't elicit much disturbance to these areas. Although the latter time interval (1976.08.16 – 1988.11.06) cumulates much more events than the previous time interval (1976.05.29 – 1976.08.16), its events is much more temporally sporadic so it has a same or even bigger Z value, imply seismicity decrease. In the area where the binary events in 1976.05.29 have impact, the z map show blue and the seismicity significantly increase. It is interesting that no red points cover region Nj, meaning that the binary events do not have transparent impact on the Nujiang fault zone though region Nj is adjacent to region Dr where the binary events happen. This may indicate that the region Nj and region Dr is not closely related during this period.

Fig.8(f) shows seismicity decrease in sporadic areas of the Sichuan-Yunnan rhombic block and its adjacent regions, while in most of the peripheral area of the Sichuan-Yunnan rhombi block, such as Nujiang fault zone, Sichuan basin, and northern part of the Sichuan-Qinghai triangle block, seismicity increases or stays unchanged.

Fig.8(g) is cover by blue Z value and blue point, illustrating large range of area start to be seismically active and indicating a forthcoming super big event which is the Mw 8.0 Wenchuan earthquake event happened in 2008.05.12.

Fig.8(h) denotes that after the Wenchuan earthquake event, large area significantly go back to seismic quiescence. From this sub-map, we can also directly find which fault zones is closely related to Longmenshan fault region. Large north-to-south area show a color of dark red, indicating the innate mechanism is north-southern oriented. Small blue-covered western domain may have a different innate mechanism.

In the Fig.8(i), the basic distance pattern performs well in the blue-covered eastern domain which surrounds the previous event (yellow diamond symbol), but the same blue-covered western domain around the middle event (yellow circle symbol) betrays the pattern. The Mw 7.3 event doesn't run out of the locally stored energy, and in the western part of the region Nj, Lc and Ys, their seismicity still keeps increasing since Fig.8(h). And some potential correlation causes the red area's seismicity to decrease.

From the 9 sub-maps, we can conclude that the spatial change of seismicity is continuous, without sudden transformation from active to inactive, and there is always a transitional gray area between the blue and warm color. This could mean a continuous geodynamic mechanism and gradual change of the stored potential energy in the eastern margin of the Qinghai-Tibetan Plateau.

Furthermore, in most of the 9 sub-maps, the latter event (yellow triangle symbol) located in the blue-covered area. It means that before big events, local seismicity will significantly increase. Moreover, if we go back to cast the locations of all yellow triangle symbols before they first show as a latter event, we can find that they generally keep showing at the blue-covered area. Hence, when big events subsequently happen, seismicity in most places oscillates from increasing to decreasing, but the epicenter of the next big event often locates where the seismicity keeps increase. From the consecutive Z value maps, we can directly focus on the always blue-covered place to forecast where next events may happen.

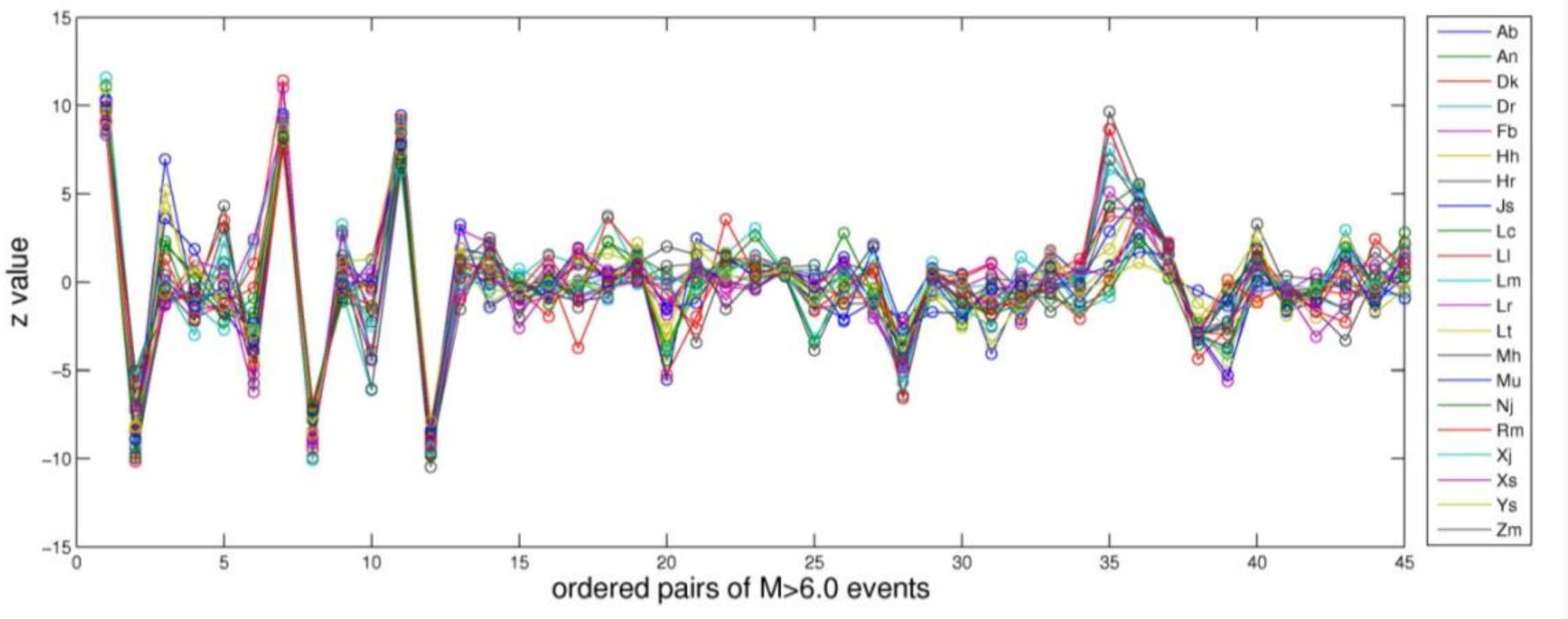

Fig.9 The Z value changes with Mw>6.0 sequence.

**5.2 Results of correlation**

Using the method in section 3.2, we calculate the respective Z value and E value sequences of all the 21 fault zones (Fig.9 ~ Fig.11), and obtain three correlation coefficients chess table (Fig.12 ~ Fig.14). Different from the time interval used in the previous section, the time intervals are divided by 47 earthquake events whose magnitude is bigger than Mw 6.0. This is because the scant 9 Mw>7.0 events cannot achieve the conventional statistical criterion of calculating the correlation, but 47 events are adequate to obtain statistical validity.

In Fig.9, we can see the Z values change with the time-ordered sequence of pairs of adjacent big events. The 21 curves of Z value display a similar tendency. Before the $13^{th}$ Z value (the seismicity rate change between 1979.03.29 - 1981.01.24 and 1981.01.24 – 1982.06.16), the general change of Z value is acute, values drastically arrive at 10 and then suddenly down to -10. This may ascribe to the deficiency of detectability of the local seismic network. As showed in Fig.10, most Mc curves are absented in the time period 1970 – 1980, due to the scant number not up to Mc's threshold 100. When the detectability is limited, seismic network tends to record the events with big magnitudes and miss the ones with small magnitudes. This causes the high seismicity rate higher and the low lower.

After the $13^{th}$ Z value, the general fluctuation of Z value starts to become small, and until the $35^{th}$ Z value (the seismicity rate change between 2007.06.03 – 2008.05.12 and 2008.05.12 – 2008.08.05), the general Z value abruptly become abnormally high, meaning the seismicity significantly decrease, which is commensurate with a theory that after a super earthquake the seismicity promptly decrease. And then the $39^{th}$ Z values go down to the nadir, meaning the most significant seismicity increase. From the $35^{th}$ and the $39^{th}$ Z value, the general change is linear and the slope is a negative constant, meaning after an abrupt seismicity decrease due to a super earthquake event, the seismicity rate keeps an unchangeable acceleration to an opposite state. Moreover, the 7 values from the $34^{th}$ to $40^{th}$ Z values almost form a standard shape of "N", which may show a particular recovery process from a super earthquake.

We used the $13^{th}$ Z value to $45^{th}$ Z value data in the Fig.9 to obtain the correlation coefficients related to seismicity rate change.

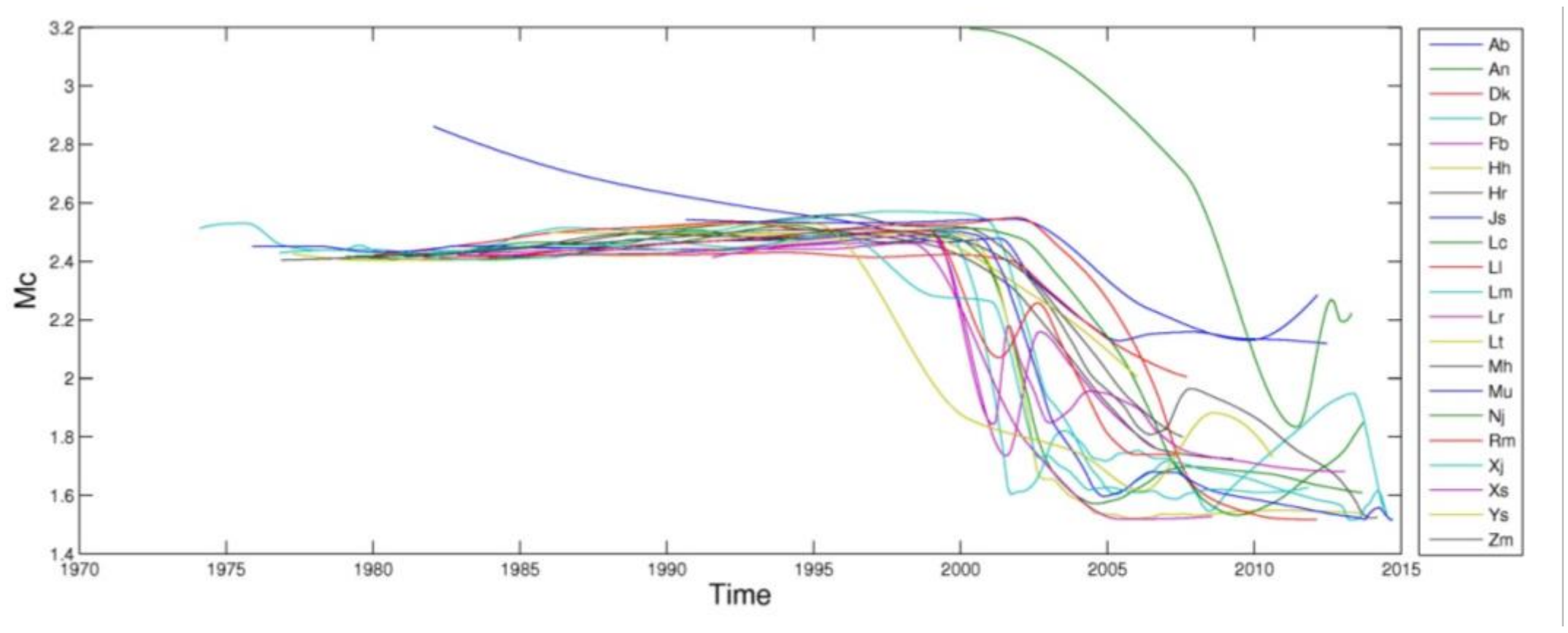

Fig.10 The Mc value changes with Mw>6.0 sequence.

Fig. 10 shows a Mc value changing with time in all the 21 fault zones. Because we set a threshold of 100 to assess the Mc, so in some time the Mc value is none. We can see that in all the fault zone the Mc value starts to decrease around 2000, which is highly possible caused by upgrading the local seismic network. We used the date in the Fig.10 to calculate the correlation coefficients related to completeness.

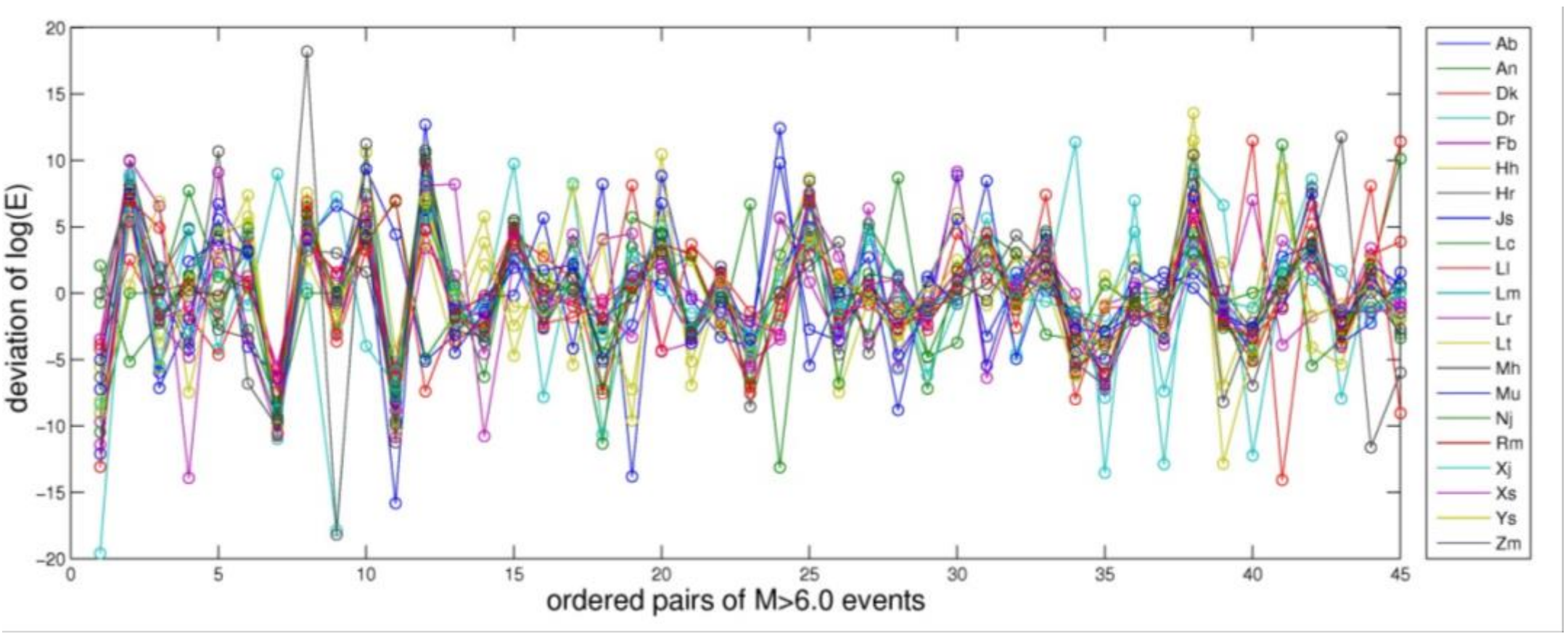

Fig.11 The E value changes with Mw>6.0 sequence.

Fig.11 show a deviation of two adjacent log(E) values varying with the temporally ordered pairs of the time intervals between two big events whose magnitude is bigger than 6.0. It shows a change of the magnitude of energy in difference fault zones, which can also be construed as an exaggerated change of earthquake magnitude. The tendencies of each curves are not so similar as ones in the Fig.9, which could conclude that when use the seismicity to access the correlation of fault zones in the eastern margin of Qinghai-Tibetan Plateau, the faults show a closer correlation in seismic frequency change than in seismic intensity change.

Unlike Z value, the E value is not significantly influenced by the absence of Mc in the sequence of $1^{st}$ - $13^{th}$ pairs of Mw>6.0 events. This is because although a seismic network with limited detectability can't completely detect the small events, it can still identify all the big events. When we calculate the energy value, the difference of energies between a big event and a small event is so huge that one can even ignore the energies of small events. Since the local seismic network can efficiently detect the relatively big events, the E value shows no patent abnormity. Hence, we will also use all the data in the Fig.11 to calculate the correlation coefficients related to energy.

Table 2 The chess table of correlation of Z values

| | Ab | An | Dk | Dr | Fb | Hh | Hr | Js | Lc | Ll | Lm | Lr | Lt | Mh | Mu | Nj | Rm | Xj | Xs | Ys | Zm |
|---|---|---|---|---|---|---|---|---|---|---|---|---|---|---|---|---|---|---|---|---|---|
| Ab | | | | | | | | | | | | | | | | | | | | | |
| An | | | | | | | | | | | | | | | | | | | | | |
| Dk | | | | | | | | | | | | | | | | | | | | | |

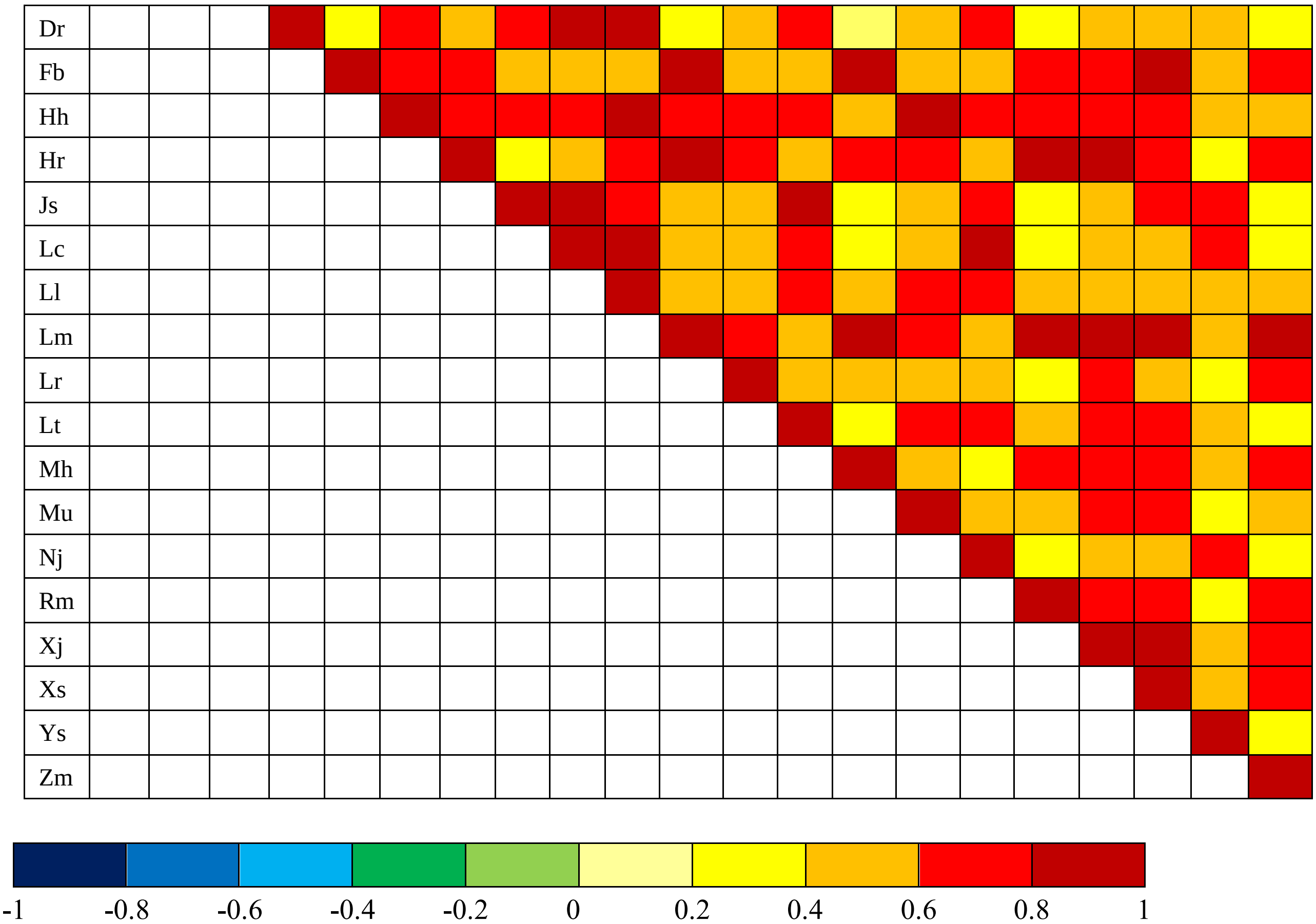

In the Table 2, we can see no cold color, which indicates that all the correlations of seismicity rate change are positive and one fault zone with increased seismicity will promote other fault zones to become more active. Moreover, the different shades of warm color show the degree of relevance. If we draw a cross with the intersection at the diagonal, we can observe that the region Dk and region Lr have no dark red correlation (0.8 – 1) with other fault zones, while the region Lm and An have respectively 7, 5 dark red correlation (0.8 - 1), the two most number. This does not only mean that the Dongkunlun fault and Longriba fault are less mutually influenced by other faults while the seismicity rate changes of Longmenshan fault and Anninghe fault are more likely to trigger and be triggered by other faults, but also indicate that the innate mechanisms are strikingly varied.

Table 3 The chess table of correlation of Mc values

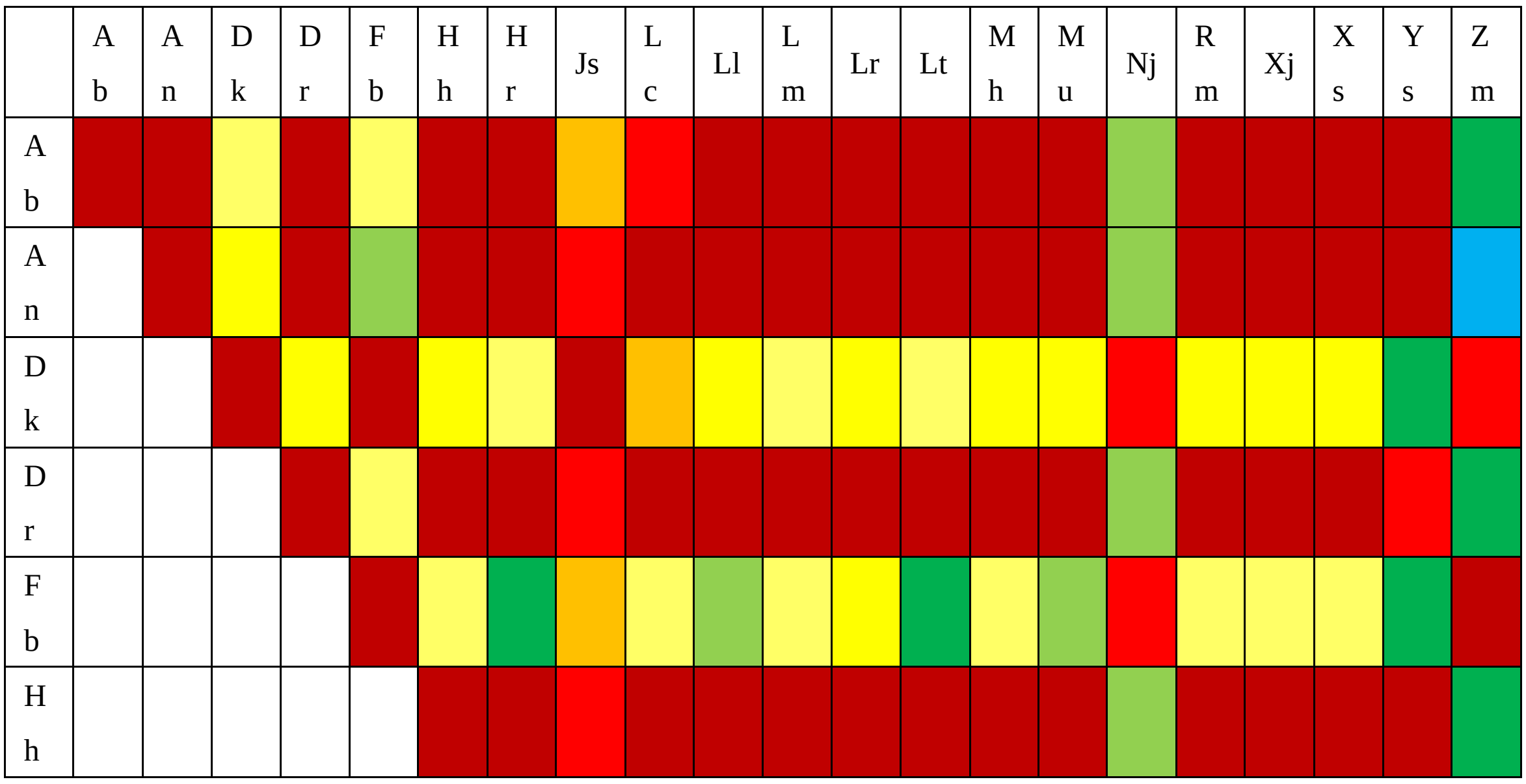

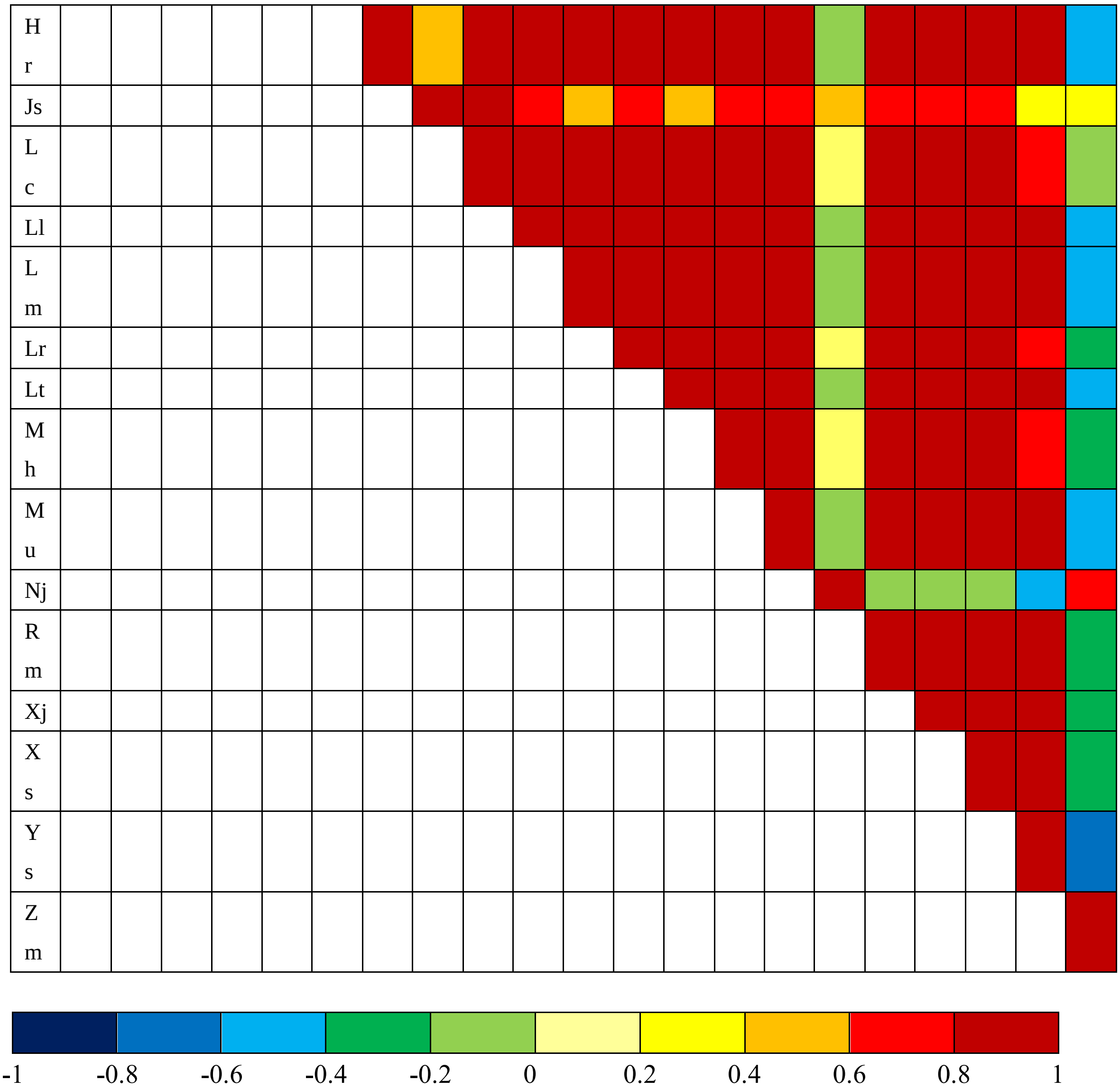

Table 3, where some lattices are not dark red, explicitly shows the spatial heterogeneous detectability of local seismic networks. The dark red lattices mean the changes of local seismic networks are almost synchronous and the relevant two fault zones have a more reliable correlation coefficients of Z value and E value. In the other way, the cold color lattices indicate the very different and even contrary changes of detectability, which could compromise that the accuracy of correlations calculated by Z value an E value. We also put a cross with intersection at the diagonal and find that the most lattices belong to region Fb, Nj and Zm are cold color. This means that the local seismic networks in these three regions need to be improved.

Table 4 visually shows the energy correlation degree between 21 fault zones.

Table 4 The chess table of correlation of E values

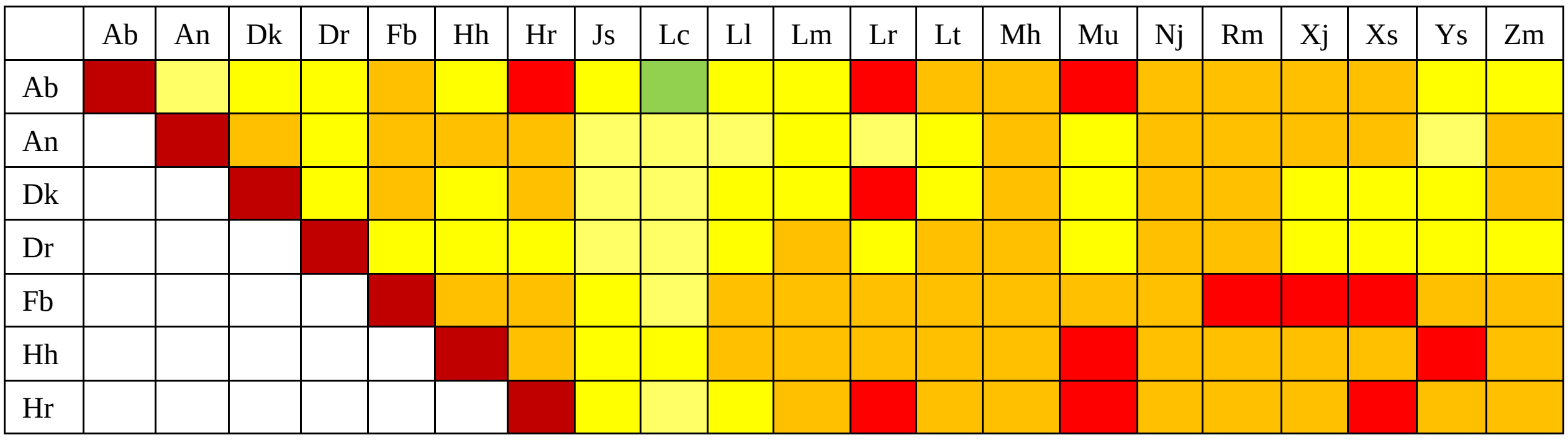

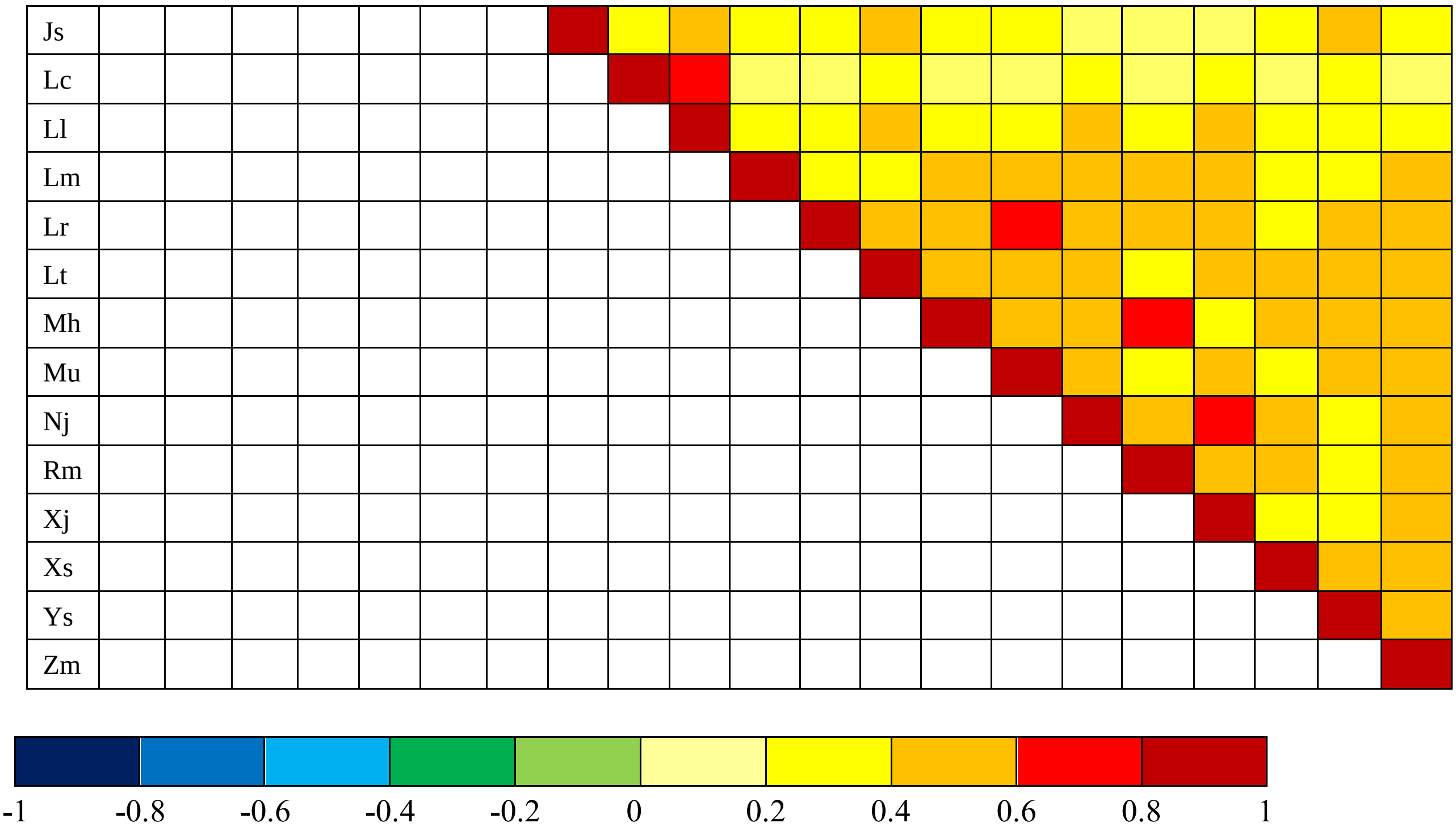

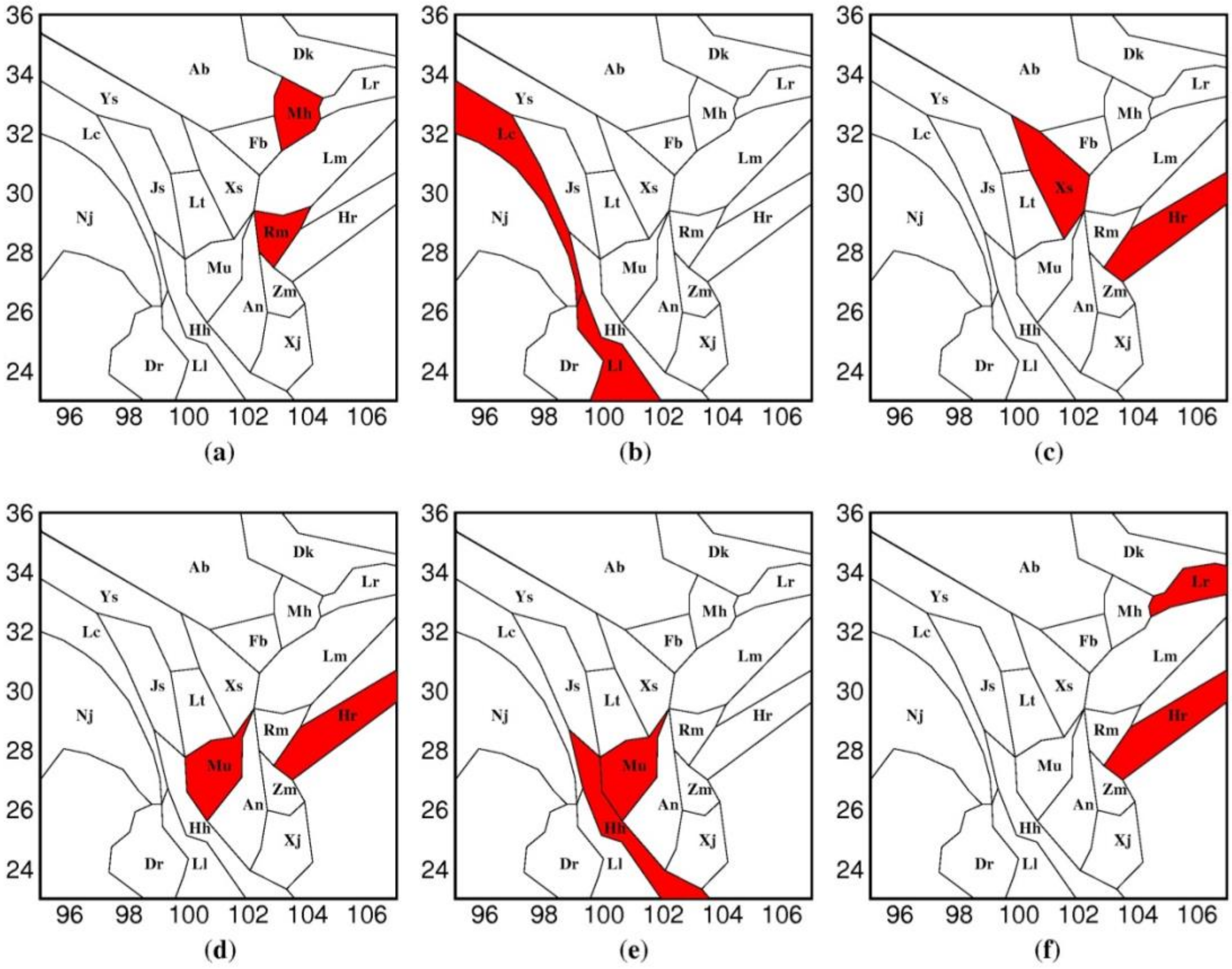

Fig.12 Six combinations with high correlation coefficients (>0.6) of both Z, Mc and E values.

## 6 Discussions

Fig.12 shows six combinations in which the relevant regions have three relatively high correlation coefficients (>0.6) of Z, Mc and E values with each other. Three types of correlation coefficients simultaneously show an affinity justify that the relevant faults stay long-term similar changes in seismic frequency, seismic intensity and completeness. So, when we access the hazard in the eastern margin of Qinghai-Tibetan Plateau, we will consider each of the six combinations as a whole, and if the seismicity of one fault zone increases, the other one's seismicity has a very definite possibility to increase too.

On the other hand, four combinations have two non-adjacent fault zones. This could show that the internal factors are superior to the external performance, such as spatial distance and length of adjacent boundaries. Furthermore, only the Lancangjiang seismic belt, composed by Lancangjiang fault and Longlin fault, is displayed, but other seismic belts do not perform an overall affinity between their own fault zones. This could mean that the spatial distribution of earthquakes can classify the consistent or continuous strikes of faults into a seismic belt, but it can't simply represent the innate correlations of the faults. Likewise, the union of faults in the same block generally can show a tendency different from other blocks, but a specific fault in a block may have a higher degree of similarity to other fault in a different block than the faults in the same block. For instance, in the Fig. 12(d), region Mu (Muli fault zone), belonging to Sichuan-Yunnan rhombic block, are closely tied with region Hr (Huarongshan fault zone) which is in Sichuan Basin.

The shared characteristic in Fig.12 (a) (b) (d) (f) is that the strikes of faults in the same combinations are parallel, while they are perpendicular in Fig.12 (c) (e). More specifically, except that region Lc and region Ll have the same left lateral faults, in other combinations with parallel strikes, the relevant faults are both reversed but with opposite trending directions, such as in Fig. 12(a) the faults in region Mh and Rm are respectively east-trending reversed and west-trending reversed, and in Fig. 12(d) region Mu has NW-trending reverse faults and region Hr SE-trending. And in combinations with perpendicular strikes, the relevant faults are reversed or strike-slip, such as in Fig.12(c) the Xianshuihe fault is left lateral slip fault while the Huarongshan fault is SE-trending reverse fault, and in Fig.12(e) the Muli fault is NW-trending reverse fault while Honghe fault is right lateral fault. This could summarize that the principle stresses of the faults in the same combinations are in the same direction.

Apparently, the property of the fault shows the mechanism, and two close-related fault regions fairly have the similar or even same mechanism. But it needs to explain that why two faults, which are in the same block and have same-directional principle stresses, don't have high correlation coefficients (>0.6) of all the three types. Even if we exclude the three fault regions with abnormal Mc values in the table 3 (Fb, Nj, Zm), many faults like Xianshuihe faults and Muli faults don't have simultaneously high correlation coefficients of Z value and E value, though they are adjacent, belong to the same Sichuan-Yunnan rhombic block and have the same NW-SE principle stresses. From table 2 and table 4, we can see that Xianshuihe faults (Xs) and Muli faults (Mu) have a high degree of Z value and a low degree of correlation with the aspect of the seismical energy changes per unit area. This indicate that even the earthquake occurrences in the Xianshuihe fault zone can forecast the frequency of earthquakes in the Muli fault zone but they are not appropriate to estimate the magnitudes of earthquakes.

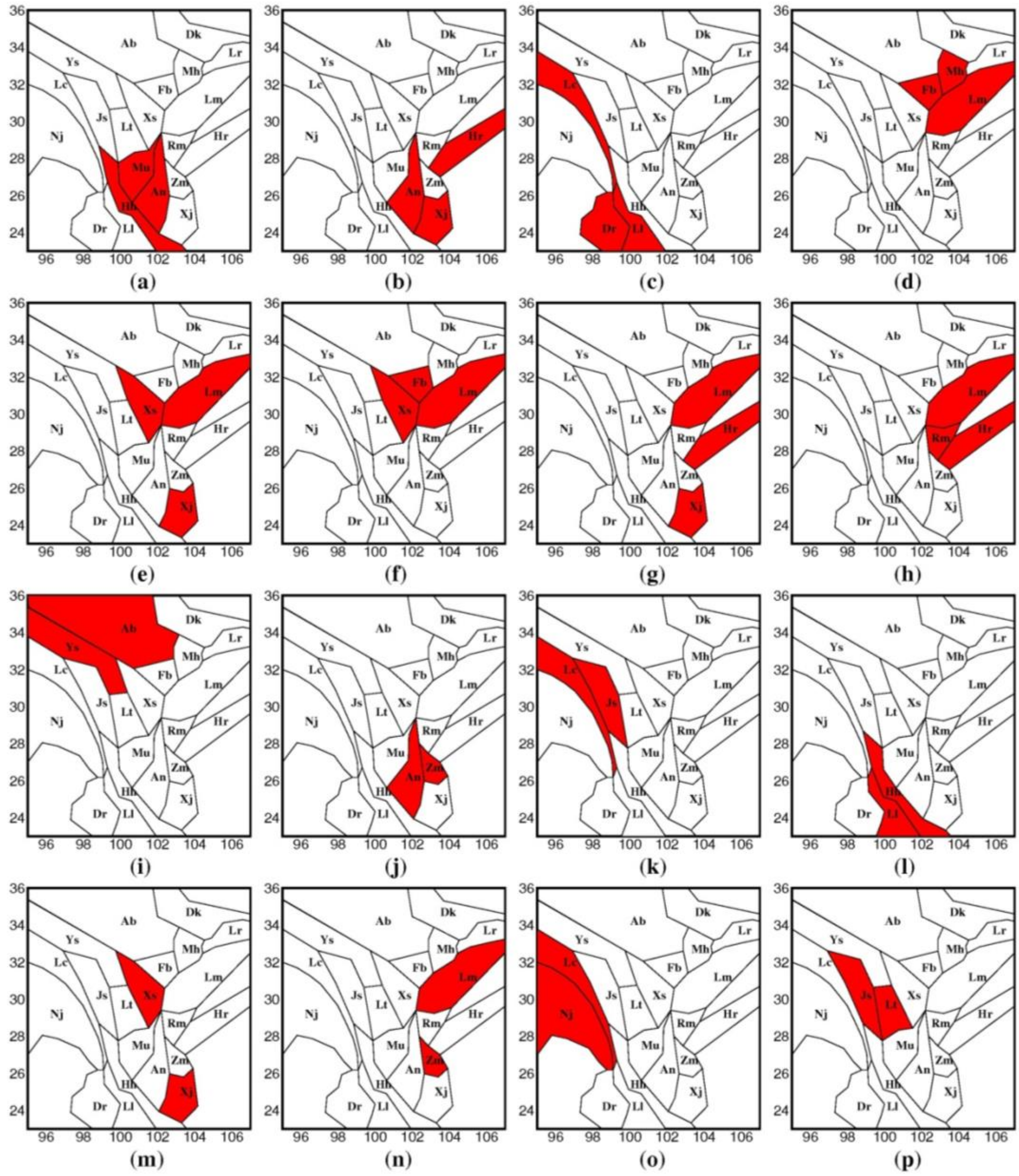

Fig.13 Sixteen combinations with correlation coefficients (>0.8) of Z values with each other.

Fig.13 shows 16 combinations of regions where the correlation coefficients are bigger than or equal to 0.8 with each other. The 16 combinations can show 16 types of close-related seismic trigger mechanisms. Moreover, the non-adjacent combinations in the 16 sub-maps can offer explanations for the cases which don't meet the distance pattern in the Fig.8. Meanwhile, the adjacent combinations can provide the distance pattern with the maximum scope of application, which is three faults zones.

Combining Fig.12 and Fig.13, we can conclude a fault relationship diagram in the eastern margin of the Qinghai-Tibet Plateau, as shown in Fig.14 (a). Fig.14 (a) shows a turntable where the order of the extent of influences from other faults clearly displays. We can further draw a more intuitive graphic on geological map, as shown in Fig.14 (b). This map tells that the famous Y-shaped area, including Longmenshan (Lm), Xianshuihe (Xs) and Anninghe (An) faults zones, and arc-shaped area, including Lancangjiang (Lc), Longlin (Ll) and Honghe (Hh) faults zones, are easily affected by other faults, because there are more faults closely related to them. In our study, we believe those easily affected faults zones are more dangerous than others.

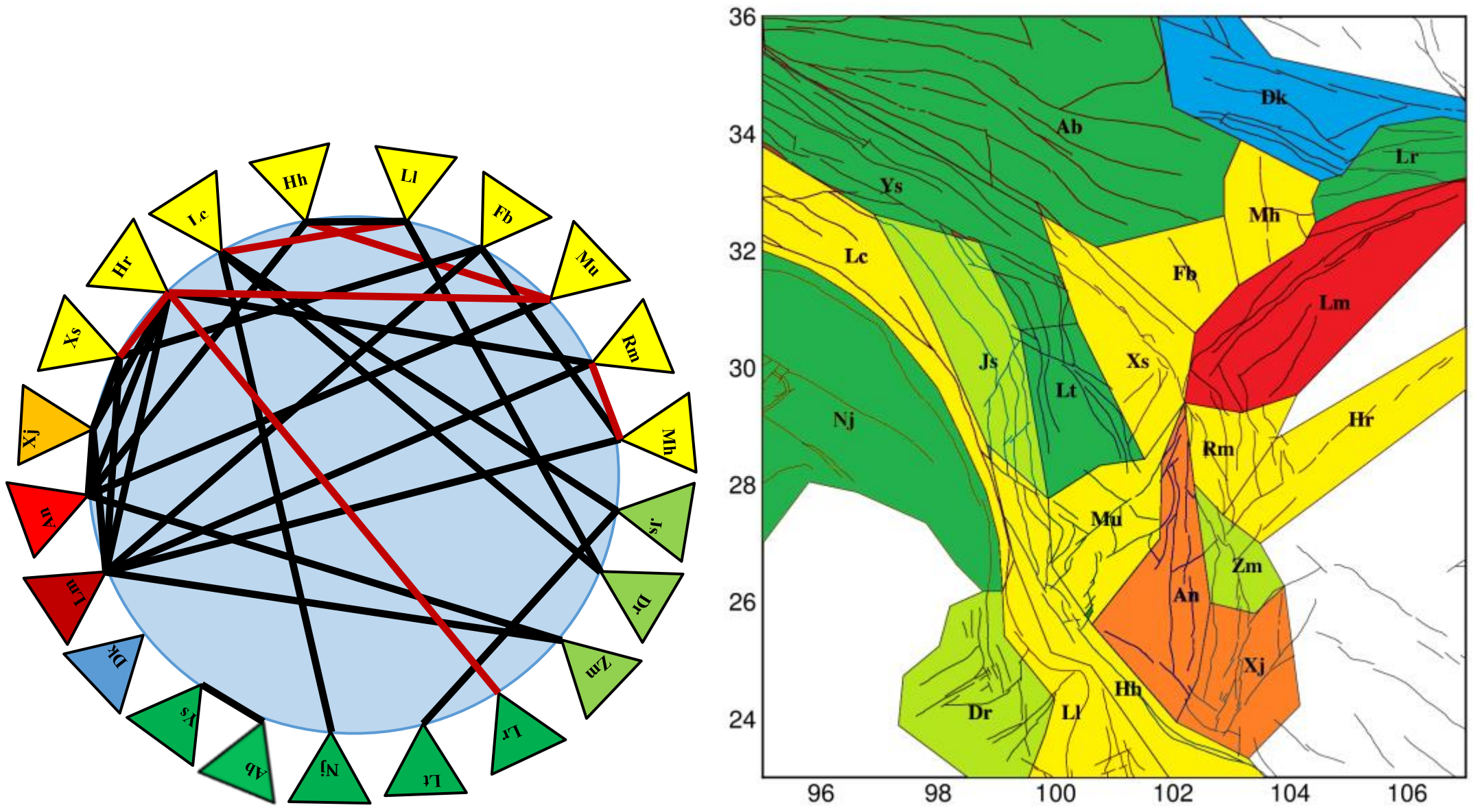

(a) (b)

Fig.14 The fault relationship diagram and its projection on the fault map. (a) the relationship diagram, the black lines represent the correlation from Fig.13 (Z correlation coefficients >0.8) and the red lines represent the correlation from Fig.12 (both Z, Mc and E correlation coefficients >0.6). The color means the number of lines: in the anti-clockwise direction, from blue (Dk) to dark red (Lm) the number increases. (b) the relationship map, the color mark is same as the one in (a).

## 7 Conclusions

In this study, we use statistical methods to calculate seismicity, energy and completeness of the contemporary earthquake catalogue in the eastern margin of Qinghai-Tibetan Plateau, and use the results to analyze the correlation of the 21 main faults zones in this area. We discover a previously unrecognized strong coupling relationship among main faults, which shows that faults like Minjiang-huya fault and Rongjiang-mabian fault can overcome the constraint of far distance and show a high closeness in the aspects of seismicity, energy and completeness. While some faults like Dayingjiang-ruili fault and Longlin fault show no highly close synchronicity of the occurrence of earthquakes nor the energy of every event, even they are adjacent. These patterns of correlation can guide researchers to better understand the deep evolution of Qinghai-Tibetan Plateau.

The seismic belts turn out to be just a group of earthquakes rather than an indicator of close internal relationship of faults. This again suggests that merely relying on spatial relation is not sufficient for understanding the deep mechanism of faults and for predicting the possible location of next big event. From the analysis of seismicity between the serial no-big-events time intervals, we could infer that the location of the area where seismicity keeps increasing comparing to every previous no-big-earthquake period might be where the next big event happens.

Since using Z-test method to analyze seismicity significantly depends on the efficient observation of small events, we have problems applying this statistical method to seismology of ancient time or even modern time, because of the limited technology of seismic network. But the period (1970-2015) we calculated is short comparing to the time of deep evolution of eastern margin of the Qinghai-Tibetan Plateau. So, to fully understand structure evolutions, a comprehensive history seismic data or a better statistical method are expected in the future studies.

## Acknowledgements

This research is supported by National Science Foundation of China (41725017, 41590864) and National Basic Research Program of China under grant number 2014CB845906. It is also partially supported by the Strategic Priority Research Program (B) of the Chinese Academy of Sciences (XDB18010202).